# A Nanocrystal Synthesis Derived Approach to Silver Bismuth Iodide Layered Double Perovskites with Aliphatic Amines: $(C_nH_{(2n+1)}NH_3)_4AgBiI_8$

*Pascal Rusch,[a,†] Ann Mary Antony,[a,b] Meenakshi Pegu,[a] Meysoun Jabrane,[a] Gabriele Saleh,[a] Arghyadeep Garai,[a,c] Aswin Asaithambi,[a] Simone Lauciello,[d] Sergio Marras,[e] Serena De Negri,[c] Pavlo Solokha,[c] Liberato Manna[a]\**

[a]Nanochemistry, Italian Institute of Technology, Via Morego 30, 16163 Genova, Italy

[b]Departimento di Fisica, Politechnico di Milano, Edificio 8, Piazza Leonardo da Vinci, 20133 Milano, Italy

[c]Dipartimento di Chimica e Chimica Industriale, Università degli Studi di Genova, Via Dodecaneso 31, Genova, 16146 Italy

[d]Electron Microscopy, Italian Institute of Technology, Via Morego 30, 16163 Genova, Italy

[e]Materials Characterization Facility, Italian Institute of Technology, Via Morego 30, 16163 Genova, Italy

**ABSTRACT**

Lead-free iodide double perovskites are an interesting class of materials since they combine a relatively low toxicity (compared to the lead counterpart) with the small bandgap typical of iodide-based perovskite structures. Their reported number is small due to their lower structural stability compared to the chloride and bromide analogues, hence their difficult synthesis. The structural constraints that limit stability, on the other hand, can be much relieved in layered, organic-inorganic perovskites. Following this line of thought, we report here a successful fast precipitation route to iodide layered $(C_nH_{(2n+1)}NH_3)_4AgBiI_8$ ($n$ = 10, 12, and 14) double perovskites that borrow concepts from the synthesis of colloidal nanocrystals. X-ray diffraction studies revealed for these compounds a monoclinic crystal structure containing edge-sharing alternated $[AgI_6]$ and $[BiI_6]$ octahedra. These materials have experimental band gaps of 2.1 eV, as also corroborated by theoretical calculations. We have also investigated their phase transitions by thermal analysis and temperature-dependent diffraction and found them to be similar to their lead-based layered perovskite counterparts.

**Introduction**

Lead halide perovskites have been widely investigated as highly promising materials, e.g. for photovoltaics,[1,2] photodetectors,[3,4] photocatalysts[5–7] and light emitters.[8–10] In these applications, the iodide perovskites have gathered considerable interest due to their smaller band gap compared to their chloride and bromide analogues, hence higher suitability for photovoltaics.[11] Unfortunately, the presence of toxic lead in these materials may represent a major concern for widespread applications.[2] One of the possible approaches to circumvent this issue is the replacement of bivalent lead ions by two different cations, a monovalent cation like $Ag^+$ and a trivalent cation like $Bi^{3+}$, resulting in a double perovskite structure.[12,13] Yet, while chloride and bromide double perovskites are fairly stable and have been widely investigated, the corresponding iodide counterparts are notably less stable, so that their investigation has advanced much less. For example, $Cs_2AgBiI_6$ is thermodynamically unstable and decomposes into $CsAg_2I_3$ and $Cs_3Bi_2I_9$.[14] On the other hand, the tight structural constraints imposed on the 3D perovskite structure are considerably relaxed in the case of 2D, layered perovskites (LPs).[15,16] This has indeed led to the emergence of LPs with Ruddlesden-Popper[17] and Dion-Jacobson[18] architectures. In the context of our work, a silver bismuth layered double perovskite (LDP) combines the replacement of $Pb^{2+}$ ions with the reduction in dimensionality, leading at once to less toxic and more structurally stable materials. Within these LDP materials, the ones incorporating iodide as anions are those that would benefit the most in terms of structural stability deriving from such reduction in dimensionality. In addition, and similarly to their lead-based counterparts, iodide-based perovskites would have smaller band gaps, which, in principle, would make them better suited for photovoltaic applications.[19,20]

There are several examples in the literature of lead iodide LPs[21–23] and silver bismuth bromide LDPs[24] in which the organic cations are the protonated versions of simple aliphatic amines, i.e. $CH_3$-$(CH_2)_n$-$NH_2$. For lead iodide LPs, a wide range of linear chain amine-based has been thoroughly synthesized and investigated regarding their temperature-dependent phase transitions,[25–27] while, surprisingly, there are no reports on the use of these simple aliphatic amines and their respective ammonium ions as organic spacers in iodide LDPs. To date, the established strategies to prepare silver bismuth iodide LDPs have been limited to the use of templating organic ammonium ions as spacers that strongly interact with each other through π-π-stacking of aromatic groups or through hydrogen bonds with heteroatoms to induce the formation of a layered structure.[20,28] The assumption in these approaches has been that the templating effect of these organic spacer ions is aiding the formation of such layered structures, as initially shown for vacancy-ordered metal iodide LPs.[29] As a consequence of the alleged need for templating spacer molecules, the number of lead-free iodide LPs has remained relatively small to date, with all syntheses requiring the use of organic cations that include in their structure heteroatoms,[28,30–33] aromatic functionalities[34,35] or both[20,28,30,36] (these are summarized in **Table S1**).

In a recent work, a combination of high-throughput synthesis with machine learning algorithms has been used to screen several amines deemed favorable for the synthesis of iodide LDPs and has thus revealed numerous possible favorable combinations. Consistent with the results mentioned above, that work concluded that the use of "bulky", structure-templating organic cations is necessary for the formation of iodide LDPs.[37] Strikingly, none of these studies has included linear chain aliphatic amines, hence leaving the impression that these molecules are not amenable to LDPs.

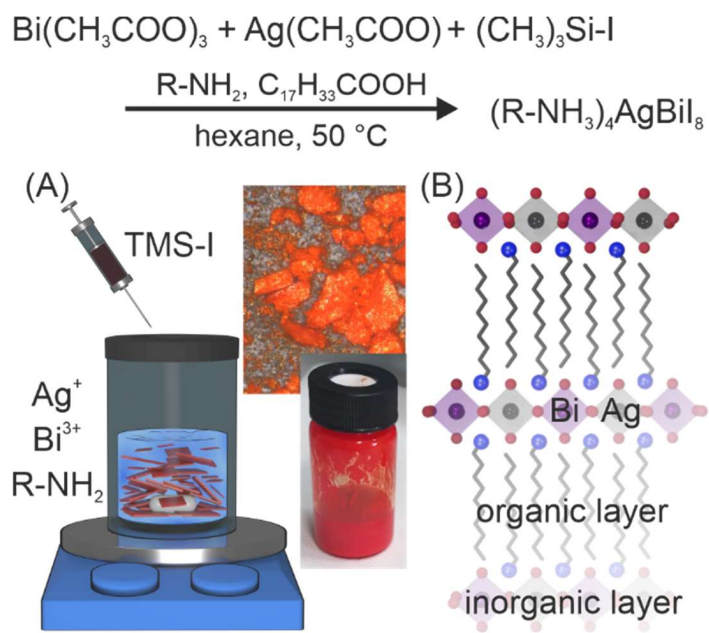

**Figure 1.** Rapid precipitation synthesis of LDP samples. Schematic depiction (A) of the rapid precipitation method used in this work to synthesize layered silver bismuth double perovskites and photographs of the synthesized microcrystals dispersed in solvent (vial) after reaction and as dried powder, and sketch (B) of their structure.

In the present work, we overcome this apparent limitation as we were able to synthesize iodide LDPs using simple aliphatic amines of various lengths. This was possible by devising an approach derived from the *hot-injection* synthesis,[38] prevalent in the synthesis of colloidal nanocrystals. Similar to the case of nanocrystal synthesis, we rely on the fast injection of a highly reactive anion precursor, which triggers the rapid nucleation and growth of layered silver bismuth iodide crystals. When using the linear aliphatic amines decylamine, dodecylamine, and tetradecylamine, this process yields silver bismuth iodide LDPs (schematically shown in **Figure 1).** These structures will be labeled C10, C12, and C14, respectively. Theoretical calculations indicate that these LDPs are semiconductors with an optical absorption edge around 2.1 eV and relatively flat bands, making a distinction between direct or indirect band gap difficult. Also, similar to the Pb-based LPs, these LDPs exhibit distinct phase transitions with temperature.

**Experimental Section**

**Chemicals.** Bismuth acetate (99.99%), silver acetate (99.99%), iodotrimethylsilane (97%, stored over Cu filings), oleic acid (90%), 1-octadecene (90%), hexane (99%), decylamine (99%), dodecylamine (97%), tetradecylamine (95%), 4-fluorophenethylamine (97%), chlorotrimethylsilane (98%), bromotrimethylsilane (97%), diethylether (99.8%), Hellmanex™, acetone (99.5%), 2-propanol (99.8%), *N,N*-dimethylformamide (DMF, 99.8%) were purchased from Sigma-Aldrich and used without further purification. Silane halide precursors are stored in a $N_2$-filled glove box (iodotrimethylsilane in the glovebox freezer at -8 °C) and handled under standard air-free techniques due to the rapid reaction of silanes with moisture and oxygen. After synthesis all samples were handled and stored under ambient atmosphere.

**Preparation of the bismuth oleate solution.** Bismuth acetate (1 mmol, 386 mg) is dissolved in a mixture of 1 mL oleic acid and 1 mL 1-octadecene by heating to 120 °C under vacuum for 30 min, followed by 30 min at 150 °C under nitrogen atmosphere. The transparent, colorless solution is left to cool to room temperature and stored under nitrogen.

**Preparation of the silver bismuth iodide LDPs.** In a vial, 0.1 mmol silver acetate, 0.5 mmol of the targeted amine, and 5 mL hexane are mixed and 0.2 mL of the above-mentioned bismuth oleate solution is added (corresponding to 0.1 mmol Bi). The reaction mixture is stirred intensely while injecting 0.1 mL iodotrimethylsilane with a syringe (or the corresponding other halide trimethylsilane). An immediate precipitation of a deep red solid can be observed. This solid is washed three times by centrifugation at 4000 rpm and redispersion in hexane using a vortexer. The solid is stored under hexane in ambient conditions. For recrystallization highly concentrated solutions were required; this synthesis procedure could be scaled up by a factor of 4 to produce ca. 800 mg of product in a single synthesis.

For recrystallization acetone is added to the hexane dispersion drop-wise until dissolution of the precipitate to form a clear, deep orange solution. The acetone-hexane solvent mix is allowed to evaporate slowly through a perforated parafilm in a saturated hexane atmosphere.

**Preparation of thin films.** Glass slide substrates for spin-coating were cleaned before deposition by consecutive immersion and ultrasonication for 20 min each in Hellmanex™ solution, MilliQ water, acetone and isopropanol. The substrates were dried under a nitrogen flow. To prepare thin films of the double perovskite samples the dispersions were centrifuged again and the supernatant discarded. The powder sample was washed two additional times with diethyl ether and dried completely. 400 mg of the resulting powder was dissolved in 0.2 mL DMF to result in a ca. 0.2M concentration. These solutions were used for spin-coating the material on the substrates mentioned above by a two-step process of 10 s at 1000 rpm and 50 s at 10000 rpm, depositing 40 μL of solution.

**Optical characterization.** Absorption spectroscopy of spin-coated sample films on glass slides was performed using an Agilent Cary5000 spectrometer equipped with a DRA-2500 integrating sphere. The samples were mounted in a slide holder in center position with an incident angle of 10°. Low temperature photoluminescence spectroscopy measurements were performed using a Renishaw inVia Reflex Raman microscope with 457 nm laser excitation focused on the sample via a 50X long working distance objective lens. The PL light is collected via the same objective, and a 600g/mm coarse grating was used for the PL measurements. An excitation power of ~ 1 mW was used. The sample was kept in Linkam flow cryostat chamber.

**Electron Microscopy.** Scanning electron microscopy images of the sample powders and crystals were collected using a Zeiss GeminiSEM 560 which was also used for elemental analysis by energy dispersive X-ray spectroscopy. The powder samples were dropcast onto Si wafers as diluted dispersions in hexane and dried under ambient conditions.

**Optical microscopy.** Light microscope images of the powders and crystals were acquired using a ZETA optical profilometer Zeta-20.

**X-ray diffraction (XRD).** Powder X-ray diffraction (p-XRD) patterns were collected using a third generation Empyrean diffractometer (Malvern-PANalytical, Westborough, MA) equipped with a 1.8kW CuKα X-ray tube operating at 45 kV and 40 mA, automated prefix iCore-dCore optical modules for the incident and diffracted beam paths, PIXcel3D area detector. Temperature dependent measurements were carried out using an Anton Paar TTK600 chamber, under $N_2$ flow (5Nl/min). Powdered samples were finely ground in an agate mortar and flattened on the holder before the measurement, while the dispersions of microcrystals were dropcast from a hexane solution. In both cases, a zero-diffraction silicon sample holder was used. HighScore Plus 5.3 software from Malvern-PANalytical was used for data analysis.

**Single-crystal X-ray diffraction.** Translucent red, plate-shaped crystals of the $(C_nH_{(2n+1)}NH_3)_4AgBiI_8$ (n=10, 12, 14) compounds were selected under a light optical microscope, glued to glass fibers and mounted on the goniometer. A three-circle Bruker D8 QUEST diffractometer using MoKα radiation (λ = 0.71076 Å) and equipped with a PHOTON III photon counting detector was used for measurements. The data collection strategies, elaborated using the APEX5 software[39] covered the reciprocal space up to a maximum θ of about 28° (resolution of *ca*. 0.75 Å). All data were integrated with SAINT V8.40B[40] and a Multi-Scan absorption correction using SADABS 2016/2 was applied.[41] The structure was solved by the intrinsic phasing method with SHELXT 2018/2 and refined by full-matrix least-squares methods against $F^2$ using SHELXL-2019/2.[42,43] All non-hydrogen atoms were refined with anisotropic displacement parameters. All C-bound hydrogen atoms were refined isotropic on calculated positions using a riding model with their $U_{iso}$ values constrained to 1.5 times the $U_{eq}$ of their pivot atoms for terminal sp$^3$ carbon atoms and 1.2 times for all other carbon atoms. Crystallographic data for the structures reported in this paper have been deposited with the Cambridge Crystallographic Data Centre[44] and they can be obtained free of charge from the Cambridge Crystallographic Data Centre via www.ccdc.cam.ac.uk/structures under the following codes: $(C_{10}H_{21}-NH_3)_4AgBiI_8$-2491406, $(C_{12}H_{25}-NH_3)_4AgBiI_8$-2366438 and $(C_{14}H_{29}-NH_3)_4AgBiI_8$-2491405. Selected crystallographic information is reported in the Supporting Material. The structures were visualized using the VESTA software package.[45]

**Thermal analysis.** Differential scanning calorimetry (DSC) was performed on a Waters-TA Instruments Discovery DSC 250 with Discovery Refrigerating Cooling System RSC90 in Aluminum non-hermetic TZero pans under nitrogen atmosphere from -90 °C to 130 °C at a rate of 5 °C/min. Before the DSC measurement, a thermogravimetry was performed to confirm no mass loss is happening in the investigated temperature range.

**Computational Details.** Density functional theory (DFT) calculations were carried out using the Vienna Ab initio Simulation Package (VASP).[46–48] The interaction between valence electrons and ionic cores was treated with the projector augmented-wave (PAW) method, and a plane-wave cutoff energy of 500 eV was used for the C10, C12, and C14 structures.[49,50] For the exchange–correlation functional, we adopted the Perdew–Burke–Ernzerhof (PBE) form of the generalized gradient approximation (GGA).[51] The atomic coordinates of all three structures were taken directly from experimental crystallographic data, and no relaxation was performed. The effect of spin–orbit coupling (SOC) was investigated separately, (re)calculating the density of states (DOS) and band gap for C12.

**Results and discussion**

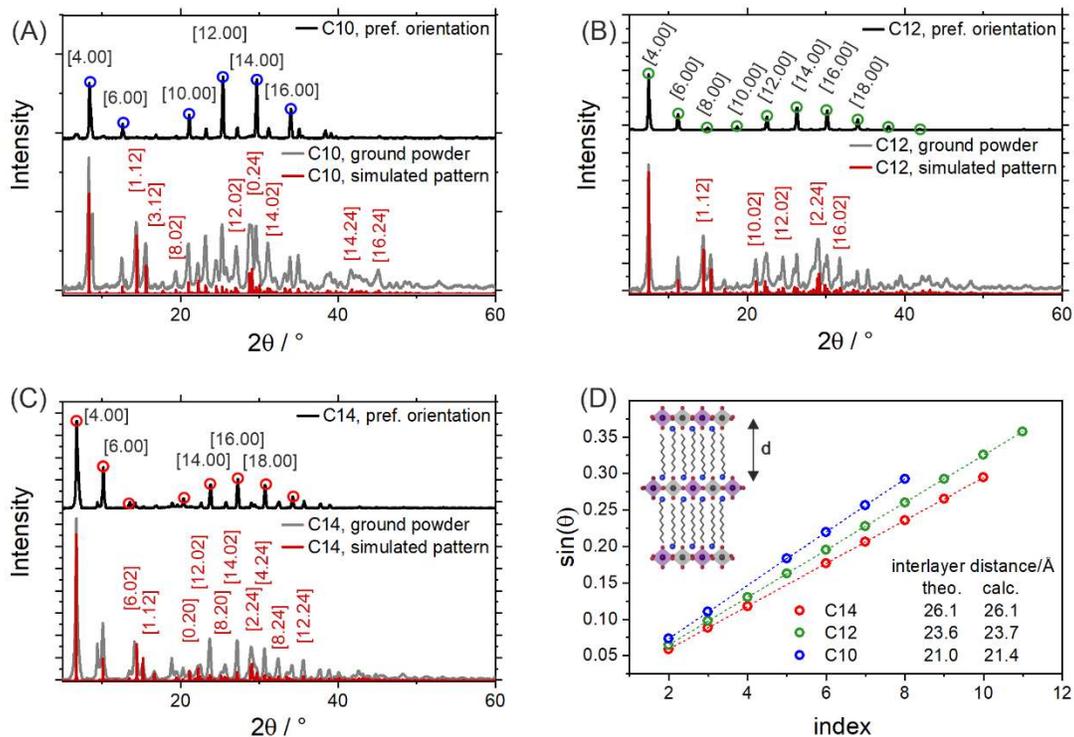

**Figure 2.** Powder diffractograms of drop-cast (A) C10, (B) C12, and (C) C14 products (black trace with the prominent [$h$.00] basal reflections marked) and comparison to diffractogram of the dried and ground powder (grey trace) and to the calculated diffraction pattern (red trace with selected additionally visible [$h.kl$] reflections marked) based on the structural model. (D) Determination of the interlayer distance (sketched in the inset) from the basal reflections.

Silver bismuth iodide LDPs were synthesized by rapid precipitation. This method consists of solubilizing the metallic ions using oleic acid and the desired amine in organic solvents, followed by the addition of trimethylsilyl iodide (schematically shown in **Figure 1A**), as detailed in the Experimental Section. This results in the immediate precipitation of a bright red or orange product (Figure 1A). The initial rapid precipitation, with the injection of the iodide precursor, aids in avoiding the formation of side products and facilitates a fast and simple synthesis of such structures.

The layered structure of the product obtained using long-chain amines (decylamine, dodecylamine, and tetradecylamine) is clearly evident from the powder X-ray diffraction (p-XRD) patterns. Dominant basal reflections ($h$00) appear with uniform spacing, corresponding to the inter-layer distance, and are indicative of a preferred orientation of the crystallites along the stacking direction (see **Figure 2**). By scanning electron microscopy (SEM), the powders appear as polydisperse microcrystals of a few μm in size (**Figure S1**). The elemental analysis of these powders by energy-dispersive X-ray spectroscopy (EDX) reveals a Ag:Bi:I ratio of ca. 1:1:8 for these samples (see **Table S2** and **Figure S5-7** for details) which matches the expected composition of a LDP. On the other hand, using shorter carbon chain amines, i.e. butylamine and hexylamine, the initial characterization did not indicate a layered structure (see **Figure S2**), and a spatially resolved elemental analysis showed a mixture of bismuth and silver iodides (see **Figure S3, S4 and Table S2**). Therefore, the work was focused on the more promising powder samples with decylamine (C10), dodecylamine (C12) and tetradecylamine (C14), on which a structural solution and further characterization were pursued. Accordingly, slow crystallization from a solvent-antisolvent mixture gave larger single crystals of a few hundred micrometers (see **Figure 2A** and **S8-10**). Therefore, it was possible to completely determine the crystal structure of the three compounds by single crystal X-ray diffraction, confirming their 2D nature: single inorganic layers of alternating corner sharing Ag-I and Bi-I octahedra are separated by organic layers of the corresponding amines (as shown in **Figure 3**).

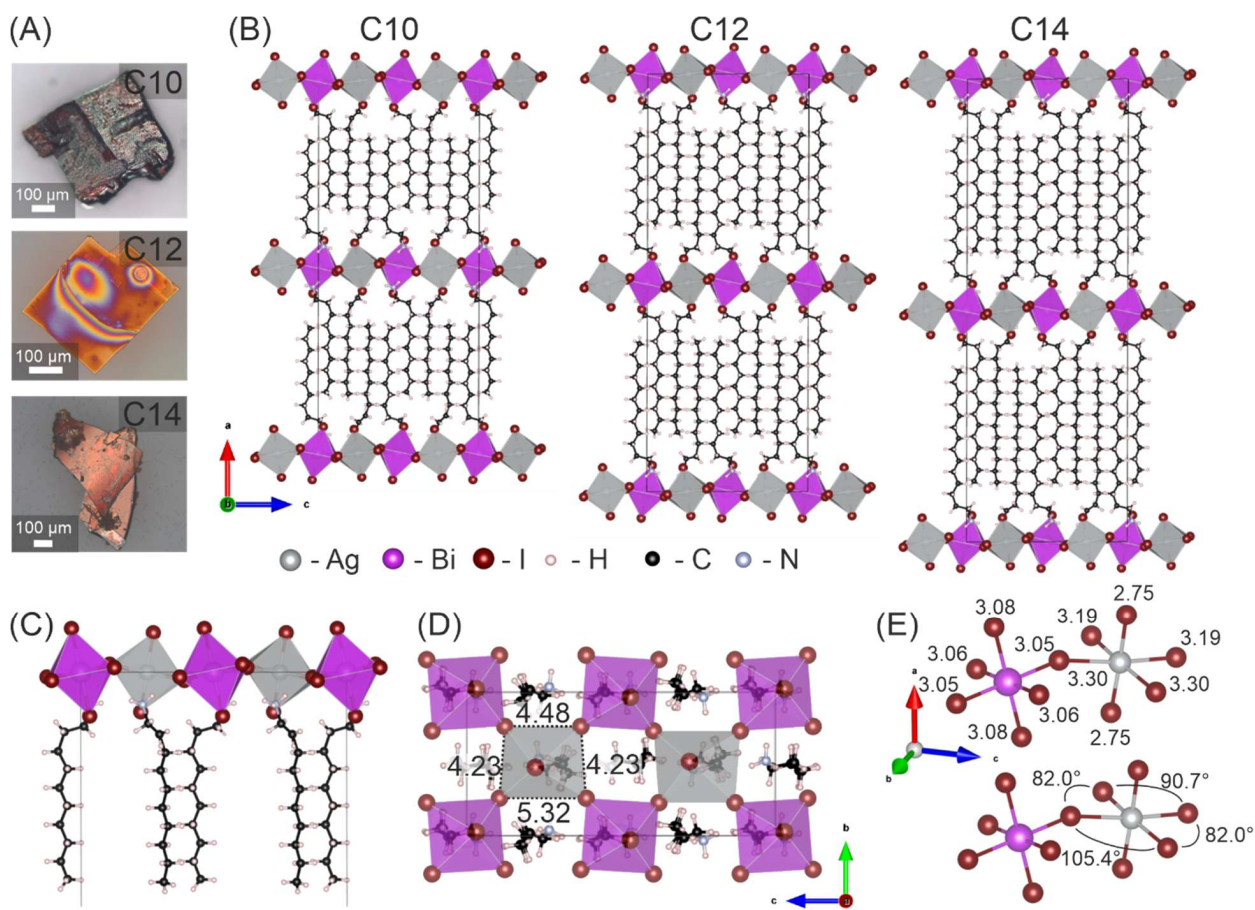

**Figure 3.** Structure of silver bismuth iodide LDPs with linear chain amines. (A) Single-crystals of samples grown using decylamine (C10), dodecylamine (C12) and tetradecylamine (C14) in optical microscopy (top to bottom) and (B, left to right) their room temperature structures viewed along the *b*-axis. (C) Enlarged image of the C10 structure highlighting the amine groups positioning to maximize their interaction with the iodide ions. (D) Crystal structure of the C10 perovskite viewed along the *a*-axis. (E) Selected distances (Å) between metal ions and coordinating iodide ions in the inorganic octahedra and angles within these octahedra for C10.

**Table 1.** Crystallographic details of the resolved structures of C10, C12, and C14 AgBiI samples at room temperature.

| CCDC number | 2491406 | 2366438 | 2491405 |
|---|---|---|---|
| Formula | $(C_{10}H_{21}$-$NH_3)_4AgBiI_8$ | $(C_{12}H_{25}$-$NH_3)_4AgBiI_8$ | $(C_{14}H_{29}$-$NH_3)_4AgBiI_8$ |
| Formula weight [gmol$^{-1}$] | 1965.25 | 2077.46 | 2189.67 |
| Crystal size [mm$^3$] | 0.03×0.08×0.1 | 0.01×0.12×0.25 | 0.04×0.19×0.21 |
| Crystal colour | translucent intense red | | |
| Crystal shape | plate like | | |
| Crystal system | monoclinic | | |
| Space group (number), Z | $C2/c$ (15), 4 | | |
| $a$ [Å] | 42.2604(13) | 47.359(4) | 52.494(3) |
| $b$ [Å] | 8.4641(3) | 8.4030(8) | 8.4011(5) |
| $c$ [Å] | 18.2342(6) | 18.2429(16) | 18.2415(10) |
| $\beta$ [°] | 90.1760(10) | 90.197(3) | 90.282(2) |
| Volume [Å$^3$] | 6522.3(4) | 7259.9(11) | 8044.6(8) |
| $\rho_{calc}$ [gcm$^{-3}$] | 2.001 | 1.901 | 1.808 |
| $\mu$ [mm$^{-1}$] | 6.807 | 6.121 | 5.529 |
| $F(000)$ | 3672 | 3928 | 4184 |
| 2θ range [°] | 4.86 to 61.05 (0.70 Å) | 4.78 to 56.70 (0.75 Å) | 4.72 to 56.85 (0.75 Å) |
| Index ranges | −60 ≤ h ≤ 60<br>−12 ≤ k ≤ 12<br>−26 ≤ l ≤ 26 | −62 ≤ h ≤ 62<br>−11 ≤ k ≤ 11<br>−23 ≤ l ≤ 24 | −70 ≤ h ≤ 69<br>−11 ≤ k ≤ 11<br>−24 ≤ l ≤ 24 |
| Reflections collected | 168657 | 3488 | 47270 |
| Independent reflections | 9950<br><br>$R_{int}$ = 0.1143<br>$R_{sigma}$ = 0.0619 | 9044<br><br>$R_{int}$ = 0.0833<br>$R_{sigma}$ = 0.0408 | 10090<br><br>$R_{int}$ = 0.1276<br>$R_{sigma}$ = 0.0689 |
| Completeness to θ = 25.243° | 99.9 | 100.0 | 99.9 |
| Data / Restraints / Parameters | 9950 / 0 / 250 | 9044 / 0 / 286 | 10090 / 0 / 322 |
| Goodness-of-fit on $F^2$ | 0.978 | 1.165 | 1.093 |
| Final $R$ indexes [$I \geq 2\sigma(I)$] | $R_1 = 0.0426$ $wR_2 = 0.0778$ | $R_1 = 0.0695$ $wR_2 = 0.1563$ | $R_1 = 0.0591$ $wR_2 = 0.1512$ |
| Final $R$ indexes [all data] | $R_1 = 0.1220$ $wR_2 = 0.1015$ | $R_1 = 0.0990$ $wR_2 = 0.1670$ | $R_1 = 0.1140$ $wR_2 = 0.1793$ |
| Largest peak/hole [eÅ$^{-3}$] | 1.43/−0.79 | 1.88/−1.37 | 2.34/−1.97 |

All compounds crystallize in the same monoclinic space group $C2/c$ (more details are listed in **Table 1**) and show a very similar architecture. This explains why the cell metrics are also similar, except for the $a$ parameter that correlates with the length of the olefin chain of the organic spacer. Accordingly, it changes from 42.2604 Å (C10) to 47.3629 Å (C12) and 52.494 Å (C14). A rough estimate for the $a$ parameter can be derived from the length of the aliphatic amine[52–54] and the average length of bismuth-iodine bonds:[55]

$$d = 2 \times (\text{Bi} - \text{I bond length}) + (\text{length of aliphatic amine with } n \text{ carbons})$$

$$d = 2 \times 3.1 \text{ Å} + (2 \text{ Å} + 1.28 \text{ Å} \times n)$$

In this case the cell contains two of these slabs and the parameter $a$ corresponds well to the theoretical assumption with C10: $2d = 2 \times 21.0 = 42.0 \approx 42.2604$, C12: $2d = 2 \times 23.6 = 47.2 \approx 47.3629$, and C14: $2d = 2 \times 26.1 = 52.2 \approx 52.494$. These values also correspond well with those extracted from p-XRD (see **Figure 2D**) on drop-cast samples. The p-XRD of the dried and ground initial powders was also measured to minimize the influence of the ($h$00) basal reflections due to preferential orientation. The resulting diffraction patterns match well with the calculated ones (see **Figure 2A-C**). Importantly, this leads us to conclude that the LDP structure does not simply form during the recrystallization but is present from the initial rapid precipitation synthesis.

In the inorganic layer, the connected Ag and Bi octahedra are tilted with respect to each other (Ag-I-Bi angle of 167.3°). The [BiI$_6$] octahedra within these layers are almost symmetrical, with all I-Bi-I angles close to 90° and only a minor difference between the Bi-I bond lengths in axial and equatorial positions. On the other hand, the [AgI$_6$] octahedra are strongly distorted. The Ag-I bond lengths are considerably longer in the equatorial direction than in the axial direction (by ca. 0.5 Å, see **Figure 3E, S11**). The two different equatorial bond lengths are adjacent: this results in the octahedral plane being a trapezoid with the sides defined by the I-I distances (as visualized in **Figure 3D** as top-down view, see also **Figure S12**). The exact bond lengths and angles differ only minutely between the C10 structure and its respective C12 and C14 analogues. Within one row of [AgI$_6$] octahedra all the mentioned trapezoids are oriented in the same direction towards the *b* axis. The next respective rows of [AgI$_6$] octahedra in *c* direction show a reversed orientation of these trapezoids (see also **Figure 3D**). Looking at the organic spacer two observations can be made: 1) they appear to occupy two different positions, one in which all carbons are fairly well localized and a second one in which the first three carbons adjacent to the amine group display a much higher flexibility in position (see **Figure S13**). This difference is likely caused by the octahedral tilt to maximize the interaction between the ammonium cation and the iodide anions. Additionally, the strong asymmetry in the Ag octahedra results in a larger space available for the amine next to the "long" sides of the trapezoid described above and a smaller space next to the "short" sides (visualized in **Figure 3D,E**); 2) considering the long, flexible carbon chain of the amines, the atomic positions in this chain are remarkably defined (see the probability ellipses in **Figure S13**). This suggests a high order within the interjecting alkyl chains. Organic moieties which interact strongly, e.g. containing pyrene groups forming π-π-stacks[35] or polar heteroatoms like 4-iodobutanamine,[33] inducing H-bond formation, have been shown to promote the formation of layered iodide double perovskite structures. Here, we assume a similar effect caused by interdigitated alkyl chains of the used aliphatic amines. While the strong ionic binding in the inorganic layer is the likely driving force behind the formation and precipitation of the product structures, the van-der-Waals interactions between the alkyl chains can serve to direct an evolution of the layered structure.

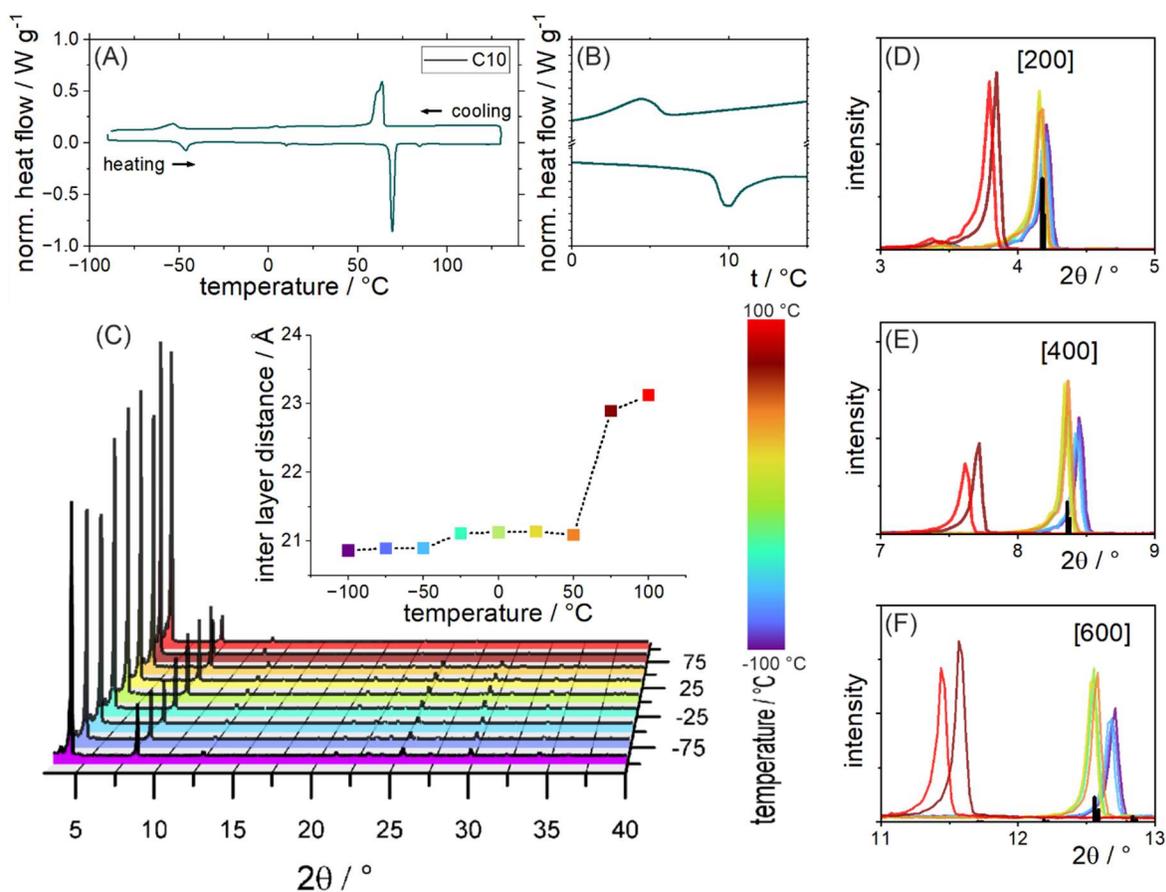

**Figure 4.** DSC of the powdered (A) C10 silver bismuth iodide LDPs with region around 10 °C shown separately (B) to visualize the small thermal event there. (C) p-XRD measured at different temperatures of the C10 sample, with the interlayer distances extracted from the [$h$00] basal reflections at different temperatures for the C10 and C12 sample. The changes at the first three

observed major reflections corresponding to (D) [200], (E) [400], and (F) [600] are shown in detail, along with the reference positions calculated from the resolved single crystal structure.

Layered lead iodides, having structures comparable to those described for the double perovskites herein, have been reported to exhibit multiple distinct phase transitions with temperature.[25–27] To investigate potentially similar behavior we performed differential scanning calorimetry (DSC) and temperature-dependent p-XRD measurements (discussed in detail in the Supporting Information and shown in **Figure 4** and **S14**). Thermal events at similar temperatures compared to the literature analogues, and a prominent change of ca. 2 Å in interlayer distance above 50 °C (C10) and above 75 °C (C12) are evident from p-XRD. These results indicate a likely similar behavior between the silver bismuth iodide LDP structures reported here and their analogue lead iodide LP structures reported by Billing and Lemmerer.[25,26] This further supports the hypothesis of the aliphatic amine exerting influence on the structure of the LDP.

To evaluate the optical properties, the powder samples were dissolved in polar solvents and spin-coated onto borosilicate glass slides (light and electron microscopy images of an exemplary film are shown in **Figure S15**). The identity of these thin films was confirmed by XRD (**Figure 5A**). It is apparent that they correspond to the respective structures observed in powder and single crystals, with an even more prominent influence of preferential orientation in the thin film, which is expected. The optical properties of the three samples, C10, C12, and C14, were remarkably similar. All three samples show an absorption edge around 580 nm (**Figure 5B**) matching their red color. The optical spectra of the three materials were displayed in a Tauc plot (see **Figure S16**) to extract their band gap yielding values of 2.01 eV (C10), 2.05 eV (C12) and 2.01 eV (C14). An additional absorption feature at higher energy is visible around 450 nm. These results are similar to the ones already reported for comparable materials,[20,28,30–32,34,37] with only a minor influence of chain length of the organic cation, which again is expected since the optical absorption of the materials is dominated by the inorganic layer of Ag and Bi octahedra. LDPs similar to the ones prepared in this work have been claimed to have a direct bandgap;[20,28] therefore, the emission properties of the LDPs synthesized by us were investigated as well. At room temperature no emission signal could be observed. When cooling the samples to liquid nitrogen temperature (below the low temperature phase transition temperature discussed above) a broad photoluminescence with maxima around 575 nm, 625 nm and 660 nm could be observed for the C10 sample (**Figure S17A**). In case of the C12 sample any emission was hardly measurable even below -100 °C (**Figure S17B**). The low intensity and broadness of the emission hint at either strong trapping inhibiting recombination or an indirect band gap.

To further elucidate the electronic properties of the materials, their band structure was modeled by Density Functional Theory simulations adopting the PBE functional (as described in the Experimental Section). The band structures of C10, C12 and C14, shown in **Figure 5C,E and S18**, are quite similar, with calculated band gaps of 2.15 eV, 2.12 eV and 2.15 eV, respectively, in fairly good agreement with our experimental data. As observed in previous studies,[19,27,29–31,34,36] the conduction band minimum (CBM) mainly originates from unoccupied Bi-$p$ and I-$p$ orbitals, while the valence band maximum (VBM) is composed of Ag-$d$ and I-$p$ states, as seen in the projected density of states (PDOS) shown in **Figure 5D**. In all three materials, the conduction band shows multiple nearly degenerate minima at different high-symmetry points in the Brillouin zone, resulting in a relatively flat CBM region (red line). This indicates a high effective mass of the excited electrons, thus suggesting poor electronic conductivity. Moreover, the flatness of the bands makes it difficult to clearly classify the band gap as direct or indirect. For example, in C14 (**Figure 5E**) the CBM and VBM are located at the V point and Γ point, respectively, but the energy difference from the closest direct transition is only 0.01 eV. This direct-indirect gap difference is lower than the room-temperature thermal energy, thus making the direct gap easily accessible.

The inclusion of spin-orbit coupling (SOC) was shown in previous studies to produce a clear splitting of the Bi 6p orbitals, particularly lowering the $6p_{1/2}$ states.[20,28] Accordingly, in our system, SOC splits the conduction band states, lifting the near-degeneracy of the Bi-p states. Notably, the character of CBM and VBM (see above) is unaffected by SOC. In particular, the conduction band remains flat near its minimum (**Figure 5F**).[20,28,31] However, we note that the introduction of SOC worsens the agreement with the experimental band gap. Indeed, given the large size of the studied systems, to simulate their band gap we exploited the well-known error compensation between the adoption of the generalized gradient approximation (PBE functional) and the neglection of SOC. Nonetheless, our investigation on the effect of SOC confirmed the electronic structure features observed adopting the PBE functional without SOC.

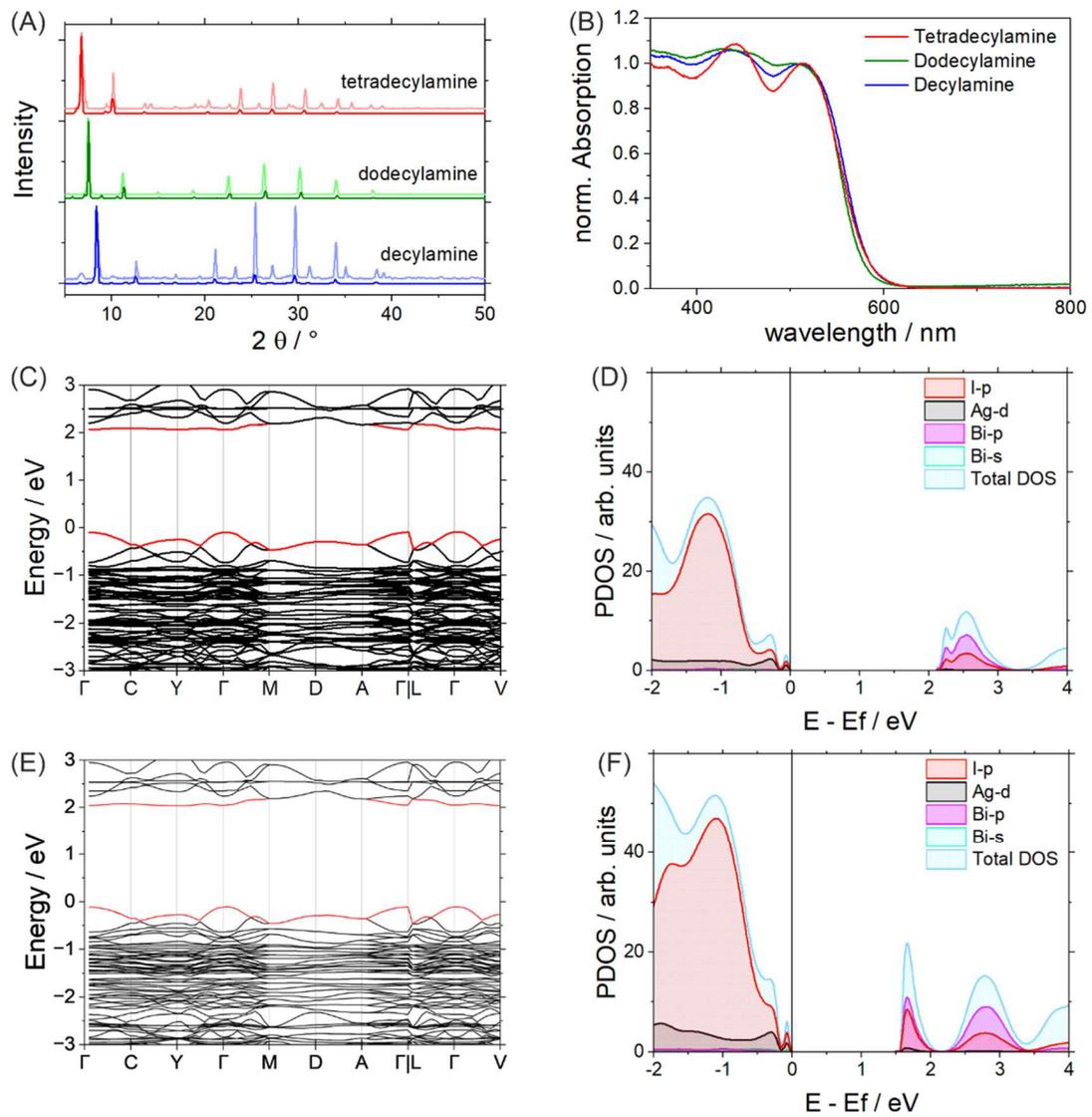

**Figure 5.** Electronic properties of the layered AgBi iodide double perovskites by optical characterization and theoretical calculations. XRD measurements (A) comparing thin film (bold trace) and dropcast powder dispersions (light trace), and room temperature absorption spectra (B) of C10, C12, and C14 $AgBiI_8$. Calculated band structure of (C) $(C_{10}H_{21}NH_3)_4AgBiI_8$ and (E) $(C_{14}H_{29}NH_3)_4AgBiI_8$. Density of states calculated without (D) spin-orbit coupling and with (F) spin-orbit coupling of $(C_{10}H_{21}NH_3)_4AgBiI_8$.

To understand the influence of our synthesis approach different parameters were varied. A controlled slow injection of iodotrimethylsilane over 5 and 60 minutes was performed to understand the formation of the product. This caused a more gradual increase in iodide concentration compared to the rapid injection. In all three reactions, independent of the injection speed, the main product is the described LDP structure (**Figure S20**). A second row of experiments was performed to understand the influence of the temperature. Here, the reaction temperature at which the iodide precursor was injected was varied between 0°C up to 50 °C. A change in injection temperature had no notable influence on the product phase (see **Figure S21**). Lastly, a comparative reaction was conducted using the established approach of slow crystallization from hydroiodic acid. Specifically, the combination of silver and bismuth iodide with decylamine in hydroiodic acid produced the same LDP structure described herein as main product (**Figure S22**). Overall, the introduced silver bismuth iodide LDPs forming largely independently of these synthesis parameters indicates them to be the thermodynamically stable product. However, when using a synthesis with slow product formation, additional reflections appear at 5-10° 2θ in the p-XRD (Figure S20,S22). This indicates the presence of contaminant side products, thus complicating product recovery and purification in these cases.

Considering the shown phase transitions close to room temperature and the general susceptibility of halide perovskite structures to moisture and ambient atmosphere the structures showed appreciable stability with the p-XRD pattern clearly matching the initial product phase. Nonetheless, a partial loss in crystallinity was apparent after ambient storage for one year (see **Figure S23**).

The generality of our synthesis approach was assessed by synthesizing the already reported 4-fluorophenethylamine (4-FPEA) silver bismuth iodide.[20] A clear match of the product using FPEA to the literature reported structure is apparent from the powder XRD (**Figure S24**). This indicates that our nanocrystal-derived synthesis approach is viable for other LDP structures as well. It was equally possible to substitute the used iodide precursor, iodotrimethylsilane, with the analogous counterparts for other halides to form the respective layered silver bismuth bromides and chlorides (**Figure S26**). As expected for halide perovskites, the introduction of other halides drastically changes the optical properties of the resulting materials, the absorption edge shifting from 580 nm (for the iodide) to 490 nm (for the bromide) to 420 nm (for the chloride) in case of the dodecylamine-containing structures investigated here (**Figure S27**). This change is also apparent in the color of the product microcrystals with the iodide being red, the bromide yellow, and the chloride colorless.

**Conclusion**

In this work we introduced an alternative synthesis route to silver bismuth iodide layered double perovskites, adapted from nanocrystal hot-injection procedures. Using this route, a set of layered silver bismuth iodides with aliphatic amines of different carbon chain lengths as organic spacers could be prepared. The powdered samples could be employed as precursors for further crystal growth to yield larger single crystals and thin films for characterization. This enabled the product structures to be solved and refined by single-crystal diffraction. The temperature-dependent characterizations revealed strong similarities between the layered double perovskites discussed here and previously reported layered perovskites containing the same amine. This reinforces the hypothesis of aliphatic amines acting as strong structure-directing agents in this case. While for structure and phase transitions an influence by the layers of interdigitated organic amines is observable, from calculations and optical spectroscopy their optical and electronic properties appear to be solely dependent on the inorganic layer. This fast injection synthesis was finally expanded to yield other literature-known similar compounds as well as the additional halides (Br and Cl) demonstrating a broader applicability of this approach.

ASSOCIATED CONTENT

**Supporting Information.**

Elemental characterization, additional structural details, temperature-dependent measurements for C12, low-temperature PL measurements of C10, C12, and C14, characterization of (4-FPEA) silver bismuth iodide, C12 silver bismuth bromide, and C12 silver bismuth chloride (PDF).

CIFs with deposition numbers 2491406 (C10), 2366438 (C12) and 2491405 (C14) contain the supplementary crystallographic data for this paper. These data can be obtained free of charge via the joint Cambridge Crystallographic Data Centre (CCDC) and Fachinformationszentrum Karlsruhe Access Structures service.

AUTHOR INFORMATION

**Corresponding Author**

*Liberato Manna, Nanochemistry Group, Italian Institute of Technology, Via Morego 30, 16163 Genova, Italy, liberato.manna@iit.it

**Present Addresses**

†Pascal Rusch, Department of Chemistry, University of Basel, Mattenstrasse 22, 4058 Basel, Switzerland

**Author Contributions**

The manuscript was written through contributions of all authors. All authors have given approval to the final version of the manuscript.

**Funding Sources**

P.R., M.P., and L.M. acknowledge funding from the Project IEMAP (Italian Energy Materials Acceleration Platform) within the Italian Research Program ENEA-MASE (Ministero dell'Ambiente e della Sicurezza Energetica) 2021−2024 "Mission Innovation" (Agreement 21A033302 GU n. 133/5-6-2021). A.M.A., M.J., A.A., and L.M. acknowledge funding from European Research Council through the ERC Advanced Grant NEHA (grant agreement n. 101095974). The computing resources and the related technical support used for this work have been provided by CRESCO/ENEAGRID High Performance Computing infrastructure and its staff.[56] CRESCO/ENEAGRID High Performance Computing infrastructure is funded by ENEA, the Italian National Agency for New Technologies, Energy and Sustainable Economic Development and by Italian and European research programs, see http://www.cresco.enea.it/english for information.

**Notes**

Any additional relevant notes should be placed here.

**REFERENCES**

(1) Kojima, A.; Teshima, K.; Shirai, Y.; Miyasaka, T. Organometal Halide Perovskites as Visible-Light Sensitizers for Photovoltaic Cells. *J. Am. Chem. Soc.* **2009**, *131* (17), 6050–6051. https://doi.org/10.1021/ja809598r.

(2) Grätzel, M. The Light and Shade of Perovskite Solar Cells. *Nature Mater* **2014**, *13* (9), 838–842. https://doi.org/10.1038/nmat4065.
(3) Wang, H.-P.; Li, S.; Liu, X.; Shi, Z.; Fang, X.; He, J.-H. Low-Dimensional Metal Halide Perovskite Photodetectors. *Advanced Materials* **2021**, *33* (7), 2003309. https://doi.org/10.1002/adma.202003309.
(4) Zheng, D.; Pauporté, T. Advances in Optical Imaging and Optical Communications Based on High-Quality Halide Perovskite Photodetectors. *Advanced Functional Materials* **2024**, *34* (11), 2311205. https://doi.org/10.1002/adfm.202311205.
(5) Zhu, X.; Lin, Y.; San Martin, J.; Sun, Y.; Zhu, D.; Yan, Y. Lead Halide Perovskites for Photocatalytic Organic Synthesis. *Nat Commun* **2019**, *10* (1), 2843. https://doi.org/10.1038/s41467-019-10634-x.
(6) Temerov, F.; Baghdadi, Y.; Rattner, E.; Eslava, S. A Review on Halide Perovskite-Based Photocatalysts: Key Factors and Challenges. *ACS Appl. Energy Mater.* **2022**, *5* (12), 14605–14637. https://doi.org/10.1021/acsaem.2c02680.
(7) Huang, Y.; Yu, J.; Wu, Z.; Li, B.; Li, M. All-Inorganic Lead Halide Perovskites for Photocatalysis: A Review. *RSC Adv.* **2024**, *14* (7), 4946–4965. https://doi.org/10.1039/D3RA07998H.
(8) Tan, Z.-K.; Moghaddam, R. S.; Lai, M. L.; Docampo, P.; Higler, R.; Deschler, F.; Price, M.; Sadhanala, A.; Pazos, L. M.; Credgington, D.; Hanusch, F.; Bein, T.; Snaith, H. J.; Friend, R. H. Bright Light-Emitting Diodes Based on Organometal Halide Perovskite. *Nature Nanotech* **2014**, *9* (9), 687–692. https://doi.org/10.1038/nnano.2014.149.
(9) Kim, J. S.; Heo, J.-M.; Park, G.-S.; Woo, S.-J.; Cho, C.; Yun, H. J.; Kim, D.-H.; Park, J.; Lee, S.-C.; Park, S.-H.; Yoon, E.; Greenham, N. C.; Lee, T.-W. Ultra-Bright, Efficient and Stable Perovskite Light-Emitting Diodes. *Nature* **2022**, *611* (7937), 688–694. https://doi.org/10.1038/s41586-022-05304-w.
(10) Kong, L.; Sun, Y.; Zhao, B.; Ji, K.; Feng, J.; Dong, J.; Wang, Y.; Liu, Z.; Maqbool, S.; Li, Y.; Yang, Y.; Dai, L.; Lee, W.; Cho, C.; Stranks, S. D.; Friend, R. H.; Wang, N.; Greenham, N. C.; Yang, X. Fabrication of Red-Emitting Perovskite LEDs by Stabilizing Their Octahedral Structure. *Nature* **2024**, *631* (8019), 73–79. https://doi.org/10.1038/s41586-024-07531-9.
(11) Kojima, A.; Teshima, K.; Shirai, Y.; Miyasaka, T. *Organometal Halide Perovskites as Visible-Light Sensitizers for Photovoltaic Cells*. ACS Publications. https://doi.org/10.1021/ja809598r.
(12) Zhao, X.-G.; Yang, J.-H.; Fu, Y.; Yang, D.; Xu, Q.; Yu, L.; Wei, S.-H.; Zhang, L. Design of Lead-Free Inorganic Halide Perovskites for Solar Cells via Cation-Transmutation. *J. Am. Chem. Soc.* **2017**, *139* (7), 2630–2638. https://doi.org/10.1021/jacs.6b09645.
(13) Wu, C.; Zhang, Q.; Liu, Y.; Luo, W.; Guo, X.; Huang, Z.; Ting, H.; Sun, W.; Zhong, X.; Wei, S.; Wang, S.; Chen, Z.; Xiao, L. The Dawn of Lead-Free Perovskite Solar Cell: Highly Stable Double Perovskite Cs2AgBiBr6 Film. *Advanced Science* **2018**, *5* (3), 1700759. https://doi.org/10.1002/advs.201700759.
(14) Xiao, Z.; Meng, W.; Wang, J.; Yan, Y. Thermodynamic Stability and Defect Chemistry of Bismuth-Based Lead-Free Double Perovskites. *ChemSusChem* **2016**, *9* (18), 2628–2633. https://doi.org/10.1002/cssc.201600771.
(15) Quan, L. N.; Yuan, M.; Comin, R.; Voznyy, O.; Beauregard, E. M.; Hoogland, S.; Buin, A.; Kirmani, A. R.; Zhao, K.; Amassian, A.; Kim, D. H.; Sargent, E. H. Ligand-Stabilized

Reduced-Dimensionality Perovskites. *J. Am. Chem. Soc.* **2016**, *138* (8), 2649–2655. https://doi.org/10.1021/jacs.5b11740.

(16) Vargas, B.; Rodríguez-López, G.; Solis-Ibarra, D. The Emergence of Halide Layered Double Perovskites. *ACS Energy Lett.* **2020**, *5* (11), 3591–3608. https://doi.org/10.1021/acsenergylett.0c01867.

(17) Smith, I. C.; Hoke, E. T.; Solis-Ibarra, D.; McGehee, M. D.; Karunadasa, H. I. A Layered Hybrid Perovskite Solar-Cell Absorber with Enhanced Moisture Stability. *Angewandte Chemie International Edition* **2014**, *53* (42), 11232–11235. https://doi.org/10.1002/anie.201406466.

(18) Mao, L.; Ke, W.; Pedesseau, L.; Wu, Y.; Katan, C.; Even, J.; Wasielewski, M. R.; Stoumpos, C. C.; Kanatzidis, M. G. Hybrid Dion–Jacobson 2D Lead Iodide Perovskites. *J. Am. Chem. Soc.* **2018**, *140* (10), 3775–3783. https://doi.org/10.1021/jacs.8b00542.

(19) Walsh, A. Principles of Chemical Bonding and Band Gap Engineering in Hybrid Organic–Inorganic Halide Perovskites. *J. Phys. Chem. C* **2015**, *119* (11), 5755–5760. https://doi.org/10.1021/jp512420b.

(20) Hooijer, R.; Weis, A.; Biewald, A.; Sirtl, M. T.; Malburg, J.; Holfeuer, R.; Thamm, S.; Amin, A. A. Y.; Righetto, M.; Hartschuh, A.; Herz, L. M.; Bein, T. Silver-Bismuth Based 2D Double Perovskites $(4FPEA)_4AgBiX_8$ (X = Cl, Br, I): Highly Oriented Thin Films with Large Domain Sizes and Ultrafast Charge-Carrier Localization. *Advanced Optical Materials* **2022**, *10* (14), 2200354. https://doi.org/10.1002/adom.202200354.

(21) Sheikh, T.; Shinde, A.; Mahamuni, S.; Nag, A. Dual Excitonic Emissions and Structural Phase Transition of Octylammonium Lead Iodide 2D Layered Perovskite Single Crystal. *Mater. Res. Express* **2019**, *6* (12), 124002. https://doi.org/10.1088/2053-1591/ab53a1.

(22) Ishihara, T.; Takahashi, J.; Goto, T. Optical Properties Due to Electronic Transitions in Two-Dimensional Semiconductors $(C_nH_{2n+1}NH_3)_2PbI_4$. *Phys. Rev. B* **1990**, *42* (17), 11099–11107. https://doi.org/10.1103/PhysRevB.42.11099.

(23) Ishihara, T.; Takahashi, J.; Goto, T. Exciton State in Two-Dimensional Perovskite Semiconductor $(C_{10}H_{21}NH_3)_2PbI_4$. *Solid State Communications* **1989**, *69* (9), 933–936. https://doi.org/10.1016/0038-1098(89)90935-6.

(24) Mao, L.; Teicher, S. M. L.; Stoumpos, C. C.; Kennard, R. M.; DeCrescent, R. A.; Wu, G.; Schuller, J. A.; Chabinyc, M. L.; Cheetham, A. K.; Seshadri, R. Chemical and Structural Diversity of Hybrid Layered Double Perovskite Halides. *J. Am. Chem. Soc.* **2019**, *141* (48), 19099–19109. https://doi.org/10.1021/jacs.9b09945.

(25) Billing, D. G.; Lemmerer, A. Synthesis, Characterization and Phase Transitions of the Inorganic–Organic Layered Perovskite-Type Hybrids [(CnH2n+1NH3)2PbI4] (n = 12, 14, 16 and 18). *New J. Chem.* **2008**, *32* (10), 1736. https://doi.org/10.1039/b805417g.

(26) Lemmerer, A.; Billing, D. G. Synthesis, Characterization and Phase Transitions of the Inorganic–Organic Layered Perovskite-Type Hybrids [$(C_n H_{2n+1} NH_3 )_2 PbI_4$ ], n = 7, 8, 9 and 10. *Dalton Trans.* **2012**, *41* (4), 1146–1157. https://doi.org/10.1039/C0DT01805H.

(27) Lemmerer, A. Thermochromic Phase Transitions of Long Odd-Chained Inorganic–Organic Layered Perovskite-Type Hybrids [$(C_n H_{2n+1} NH_3)_2 PbI_4$ ], *n* = 11, 13, and 15. *Inorg. Chem.* **2022**, *61* (17), 6353–6366. https://doi.org/10.1021/acs.inorgchem.1c03132.

(28) Jana, M. K.; Janke, S. M.; Dirkes, D. J.; Dovletgeldi, S.; Liu, C.; Qin, X.; Gundogdu, K.; You, W.; Blum, V.; Mitzi, D. B. Direct-Bandgap 2D Silver–Bismuth Iodide Double Perovskite: The Structure-Directing Influence of an Oligothiophene Spacer Cation. *J. Am. Chem. Soc.* **2019**, *141* (19), 7955–7964. https://doi.org/10.1021/jacs.9b02909.

(29) Mitzi, D. B. Organic−Inorganic Perovskites Containing Trivalent Metal Halide Layers: The Templating Influence of the Organic Cation Layer. *Inorg. Chem.* **2000**, *39* (26), 6107–6113. https://doi.org/10.1021/ic000794i.

(30) Li, X.; Traoré, B.; Kepenekian, M.; Li, L.; Stoumpos, C. C.; Guo, P.; Even, J.; Katan, C.; Kanatzidis, M. G. Bismuth/Silver-Based Two-Dimensional Iodide Double and One-Dimensional Bi Perovskites: Interplay between Structural and Electronic Dimensions. *Chem. Mater.* **2021**, *33* (15), 6206–6216. https://doi.org/10.1021/acs.chemmater.1c01952.

(31) Yao, Y.; Kou, B.; Peng, Y.; Wu, Z.; Li, L.; Wang, S.; Zhang, X.; Liu, X.; Luo, J. $(C_3H_9NI)_4AgBiI_8$: A Direct-Bandgap Layered Double Perovskite Based on a Short-Chain Spacer Cation for Light Absorption. *Chem. Commun.* **2020**, *56* (21), 3206–3209. https://doi.org/10.1039/C9CC07796K.

(32) Bi, L.-Y.; Hu, Y.-Q.; Li, M.-Q.; Hu, T.-L.; Zhang, H.-L.; Yin, X.-T.; Que, W.-X.; Lassoued, M. S.; Zheng, Y.-Z. Two-Dimensional Lead-Free Iodide-Based Hybrid Double Perovskites: Crystal Growth, Thin-Film Preparation and Photocurrent Responses. *J. Mater. Chem. A* **2019**, *7* (34), 19662–19667. https://doi.org/10.1039/C9TA04325J.

(33) Xu, Z.; Wu, H.; Li, D.; Wu, W.; Li, L.; Luo, J. A Lead-Free I-Based Hybrid Double Perovskite $(I-C_4H_8NH_3)_4AgBiI_8$ for X-Ray Detection. *J. Mater. Chem. C* **2021**, *9* (38), 13157–13161. https://doi.org/10.1039/D1TC03412J.

(34) Fu, D.; Wu, S.; Liu, Y.; Yao, Y.; He, Y.; Zhang, X.-M. A Lead-Free Layered Dion–Jacobson Hybrid Double Perovskite Constructed by an Aromatic Diammonium Cation. *Inorg. Chem. Front.* **2021**, *8* (14), 3576–3580. https://doi.org/10.1039/D1QI00219H.

(35) Hooijer, R.; Wang, S.; Biewald, A.; Eckel, C.; Righetto, M.; Chen, M.; Xu, Z.; Blätte, D.; Han, D.; Ebert, H.; Herz, L. M.; Weitz, R. T.; Hartschuh, A.; Bein, T. Overcoming Intrinsic Quantum Confinement and Ultrafast Self-Trapping in Ag–Bi–I- and Cu–Bi–I-Based 2D Double Perovskites through Electroactive Cations. *J. Am. Chem. Soc.* **2024**. https://doi.org/10.1021/jacs.4c04616.

(36) Wang, C.-F.; Li, H.; Li, M.-G.; Cui, Y.; Song, X.; Wang, Q.-W.; Jiang, J.-Y.; Hua, M.-M.; Xu, Q.; Zhao, K.; Ye, H.-Y.; Zhang, Y. Centimeter-Sized Single Crystals of Two-Dimensional Hybrid Iodide Double Perovskite (4,4-Difluoropiperidinium)4AgBiI8 for High-Temperature Ferroelectricity and Efficient X-Ray Detection. *Advanced Functional Materials* **2021**, *31* (13), 2009457. https://doi.org/10.1002/adfm.202009457.

(37) Wu, Y.; Wang, C.-F.; Ju, M.-G.; Jia, Q.; Zhou, Q.; Lu, S.; Gao, X.; Zhang, Y.; Wang, J. Universal Machine Learning Aided Synthesis Approach of Two-Dimensional Perovskites in a Typical Laboratory. *Nat Commun* **2024**, *15* (1), 138. https://doi.org/10.1038/s41467-023-44236-5.

(38) de Mello Donegá, C.; Liljeroth, P.; Vanmaekelbergh, D. Physicochemical Evaluation of the Hot-Injection Method, a Synthesis Route for Monodisperse Nanocrystals. *Small* **2005**, *1* (12), 1152–1162. https://doi.org/10.1002/smll.200500239.

(39) *Bruker AXS Inc. V2021.10-0*; Bruker AXS Inc., 2022.

(40) *Saint V8.40B*; Bruker AXS Inc.: Madison, Wisconsin, USA.

(41) Krause, L.; Herbst-Irmer, R.; Sheldrick, G. M.; Stalke, D. Comparison of Silver and Molybdenum Microfocus X-Ray Sources for Single-Crystal Structure Determination. *J Appl Cryst* **2015**, *48* (1), 3–10. https://doi.org/10.1107/S1600576714022985.

(42) Sheldrick, G. M. SHELXT – Integrated Space-Group and Crystal-Structure Determination. *Acta Cryst A* **2015**, *71* (1), 3–8. https://doi.org/10.1107/S2053273314026370.

(43) Sheldrick, G. M. Crystal Structure Refinement with SHELXL. *Acta Cryst C* **2015**, *71* (1), 3–8. https://doi.org/10.1107/S2053229614024218.

(44) Groom, C. R.; Bruno, I. J.; Lightfoot, M. P.; Ward, S. C. The Cambridge Structural Database. *Acta Cryst B* **2016**, *72* (2), 171–179. https://doi.org/10.1107/S2052520616003954.

(45) Momma, K.; Izumi, F. VESTA 3 for Three-Dimensional Visualization of Crystal, Volumetric and Morphology Data. *J Appl Cryst* **2011**, *44* (6), 1272–1276. https://doi.org/10.1107/S0021889811038970.

(46) Kresse, G. Ab Initio Molecular Dynamics for Liquid Metals. *Journal of Non-Crystalline Solids* **1995**, *192–193*, 222–229. https://doi.org/10.1016/0022-3093(95)00355-X.

(47) Kresse, G.; Furthmüller, J. Efficient Iterative Schemes for Ab Initio Total-Energy Calculations Using a Plane-Wave Basis Set. *Phys. Rev. B* **1996**, *54* (16), 11169–11186. https://doi.org/10.1103/PhysRevB.54.11169.

(48) Kresse, G.; Furthmüller, J. Efficiency of Ab-Initio Total Energy Calculations for Metals and Semiconductors Using a Plane-Wave Basis Set. *Computational Materials Science* **1996**, *6* (1), 15–50. https://doi.org/10.1016/0927-0256(96)00008-0.

(49) Blöchl, P. E. Projector Augmented-Wave Method. *Phys. Rev. B* **1994**, *50* (24), 17953–17979. https://doi.org/10.1103/PhysRevB.50.17953.

(50) Kresse, G.; Joubert, D. From Ultrasoft Pseudopotentials to the Projector Augmented-Wave Method. *Phys. Rev. B* **1999**, *59* (3), 1758–1775. https://doi.org/10.1103/PhysRevB.59.1758.

(51) Perdew, J. P.; Ernzerhof, M.; Burke, K. Rationale for Mixing Exact Exchange with Density Functional Approximations. *J. Chem. Phys.* **1996**, *105* (22), 9982–9985. https://doi.org/10.1063/1.472933.

(52) Travesset, A. Soft Skyrmions, Spontaneous Valence and Selection Rules in Nanoparticle Superlattices. *ACS Nano* **2017**, *11* (6), 5375–5382. https://doi.org/10.1021/acsnano.7b02219.

(53) Hallstrom, J.; Cherniukh, I.; Zha, X.; Kovalenko, M. V.; Travesset, A. Ligand Effects in Assembly of Cubic and Spherical Nanocrystals: Applications to Packing of Perovskite Nanocubes. *ACS Nano* **2023**, *17* (8), 7219–7228. https://doi.org/10.1021/acsnano.2c10079.

(54) Zha, X.; Travesset, A. The Hard Sphere Diameter of Nanocrystals (Nanoparticles). *The Journal of Chemical Physics* **2020**, *152* (9), 094502. https://doi.org/10.1063/1.5132747.

(55) Ma, F.; Zhou, M.; Jiao, Y.; Gao, G.; Gu, Y.; Bilic, A.; Chen, Z.; Du, A. Single Layer Bismuth Iodide: Computational Exploration of Structural, Electrical, Mechanical and Optical Properties. *Sci Rep* **2015**, *5* (1), 17558. https://doi.org/10.1038/srep17558.

(56) Iannone, F.; Ambrosino, F.; Bracco, G.; De Rosa, M.; Funel, A.; Guarnieri, G.; Migliori, S.; Palombi, F.; Ponti, G.; Santomauro, G.; Procacci, P. CRESCO ENEA HPC Clusters: A Working Example of a Multifabric GPFS Spectrum Scale Layout. In *2019 International Conference on High Performance Computing & Simulation (HPCS)*; 2019; pp 1051–1052. https://doi.org/10.1109/HPCS48598.2019.9188135.

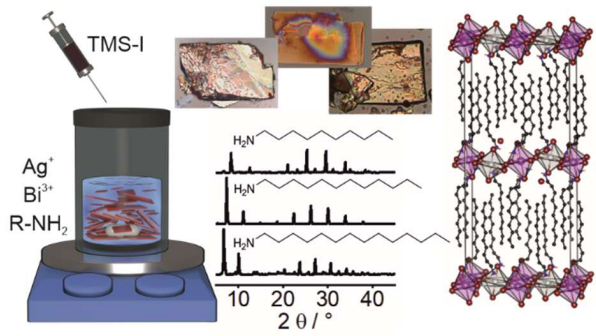

TOC Graphic 1

# Supporting Information: A Nanocrystal Synthesis Derived Approach to Silver Bismuth Iodide Layered Double Perovskites with Aliphatic Amines: $(C_nH_{(2n+1)}NH_3)_4AgBiI_8$

*Pascal Rusch,[a,†] Ann Mary Antony,[a,b] Meenakshi Pegu,[a] Meysoun Jabrane,[a] Gabriele Saleh,[a] Arghyadeep Garai,[a,c] Aswin Asaithambi,[a] Simone Lauciello,[d] Sergio Marras,[e] Serena De Negri,[c] Pavlo Solokha,[c] Liberato Manna[a]\**

[a]Nanochemistry, Italian Institute of Technology, Via Morego 30, 16163 Genova, Italy

[b]Departimento di Fisica, Politechnico di Milano, Edificio 8, Piazza Leonardo da Vinci, 20133 Milano, Italy

[c]Dipartimento di Chimica e Chimica Industriale, Università degli Studi di Genova, Via Dodecaneso 31, Genova, 16146 Italy

[d]Electron Microscopy, Italian Institute of Technology, Via Morego 30, 16163 Genova, Italy

[e]Materials Characterization Facility, Italian Institute of Technology, Via Morego 30, 16163 Genova, Italy

**Table S1.** Known examples of (Amine)$_4$AgBiI$_8$ layered double perovskites.

| Amine | | Synthesis method | Band gap | Reference |
|---|---|---|---|---|
| **4-fluorophenethylamine** | 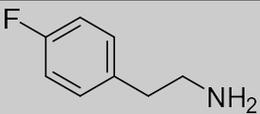 | Slow crystallization from HI | 2.16 eV | [19] |
| **5,5′-diylbis-(aminoethyl)-[2,2′-bithiophene]** | 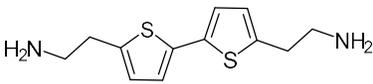 | Slow crystallization from HI | 2.0 eV | [27] |
| **4,4-difluoropiperidine** | 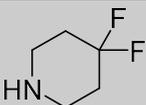 | Slow crystallization from HI | 1.93 eV | [35] |
| **4-aminomethylpiperidine** | 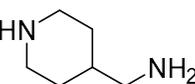 | Slow crystallization from HI | 1.27 eV | [29] |
| **4-aminomethylpyridine** | 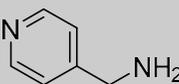 | Slow crystallization from HI | 1.44 eV | [29] |
| **4-iodobutylamine** | 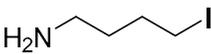 | Slow crystallization from HI | 2.0 eV | [32] |
| **3-iodopropylamine** | 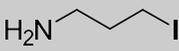 | Slow crystallization from HI | 1.87 eV | [30] |
| **3-(aminomethyl) pyridine** | 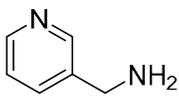 | Slow crystallization from HI | 1.86 eV | [33] |
| **naphthalene-*O*-ethylamine** | 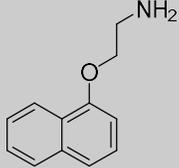 | Crystallization from polar organic solvents (DMF/DMSO) | 2.19 eV | [34] |
| naphthalene-*O*-propylamine | 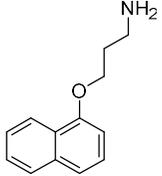 | Crystallization from polar organic solvents (DMF/DMSO) | 2.18 eV | [34] |
| **pyrene-*O*-ethylamine** | 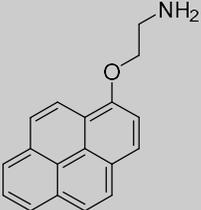 | Crystallization from polar organic solvents (DMF/DMSO) | 2.15 eV | [34] |
| pyrene-*O*-propylamine | 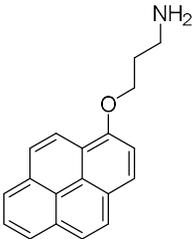 | Crystallization from polar organic solvents (DMF/DMSO) | 2.19 eV | [34] |
| **4-chlorobenzylamine** | 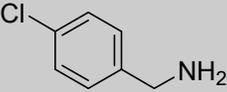 | Slow crystallization from HI | 1.89 eV | [36] |
| **3-fluoro-4-chloroanilin** | 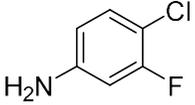 | Slow crystallization from HI | 1.85 eV | [36] |
| **3-fluoropiperidine** | 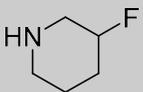 | Slow crystallization from HI | 1.94 eV | [36] |
| **4-(aminomethyl)piperidine** | 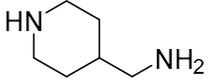 | Slow crystallization from HI | 1.88 eV | [36] |
| **histamine** | 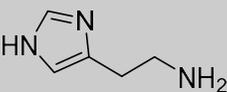 | Slow crystallization from HI | 1.84 eV | [36] |
| **4,4,4-trifluorobutan-1-amine** | 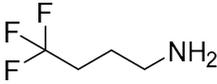 | Slow crystallization from HI | 1.97 eV | [36] |
| (S)-1-(4-chlorophenyl)ethan-1-amine | | Slow crystallization from HI | 1.93 eV | [36] |
| (S)-1-(4-bromophenyl)ethan-1-amine | | Slow crystallization from HI | 1.95 eV | [36] |
| (S)-1-(p-tolyl)ethan-1-amine | | Slow crystallization from HI | 1.93 eV | [36] |
| 3,3-difluoropyrrolidine | | Slow crystallization from HI | 1.99 eV | [36] |
| cyclohexane-1,4-diamine | | Slow crystallization from HI | 1.85 eV | [36] |
| decylamine | | Rapid precipitation | 2.1 eV | this work |
| dodecylamine | | Rapid precipitation | 2.1 eV | this work |
| tetradecylamine | | Rapid precipitation | 2.1 eV | this work |

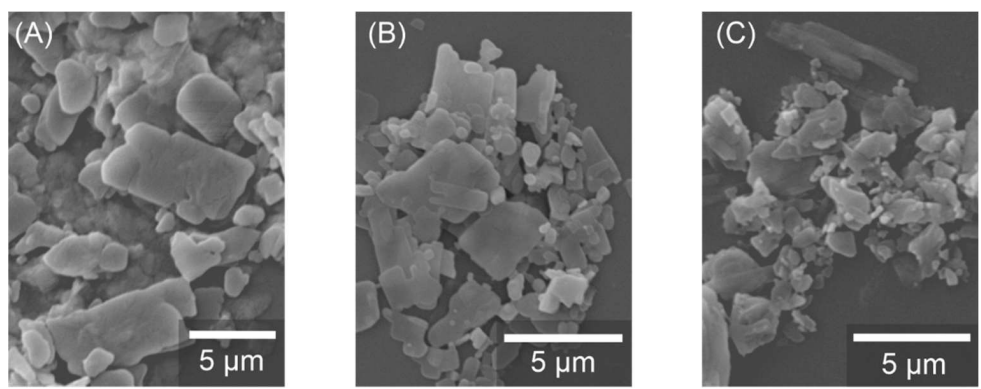

**Figure S 1.** Scanning electron microscopy images of the powdered C10, C12, and C14 samples

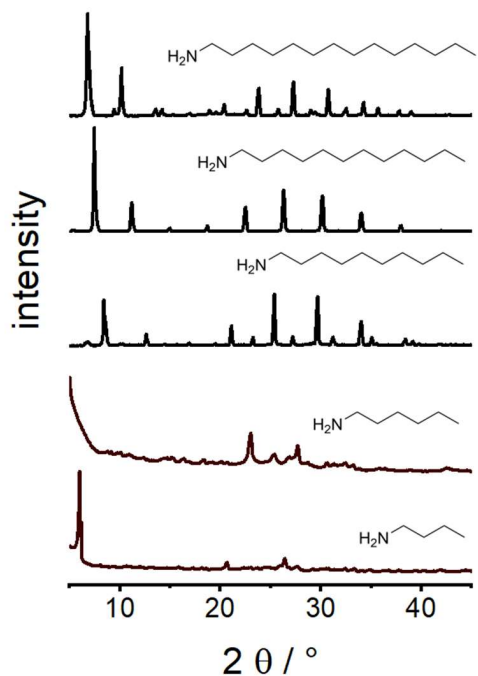

**Figure S 2.** Powder diffractograms of dropcast product dispersions using different aliphatic amines in the synthesis.

**Table S 2.** Elemental composition of product silver bismuth iodide powder using various amines as determined by SEM-EDX

| Sample (used amine) | Ag (at.%) | Bi (at.%) | I (at.%) |
|---|---|---|---|
| **Butylamine** | 7.1 (±1.9) % | 13.8 (±1.6) % | 79.2 (±0.4) % |
| **Hexylamine** | 4.1 (±0.7) % | 12.9 (±0.4) % | 82.9 (±0.9) % |
| **Decylamine** | 11.8 (±0.5) % | 8.9 (±0.1) % | 79.3 (±0.4) % |
| **Dodecylamine** | 11.3 (±0.1) % | 8.7 (±0.1) % | 80.0 (±0.1) % |
| **Tetradecylamine** | 11.3 (±0.1) % | 9.7 (±0.5) % | 79.0 (±0.6) % |
| **Fluorophenethylamine** | 9.8 (±1.2) % | 7.7 (±1.1) % | 82.6 (±2.1) % |

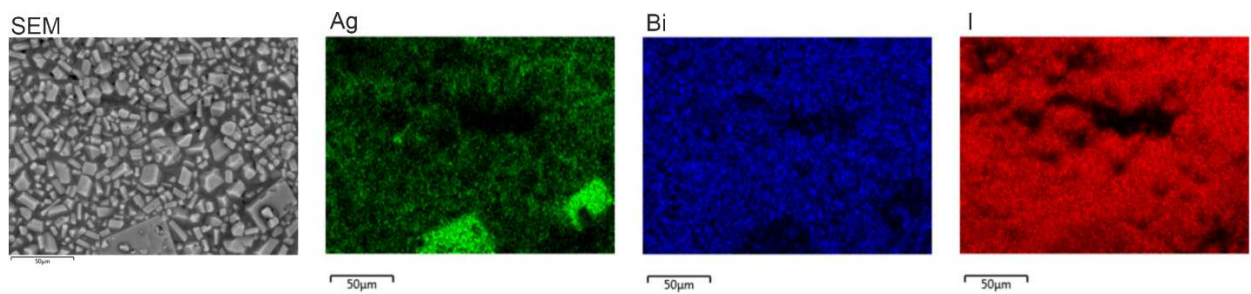

**Figure S 3.** Scanning electron microscopy overview of silver bismuth iodide synthesized using butylamine and corresponding elemental mapping of Ag, Bi and I by energy-dispersive X-ray spectroscopy.

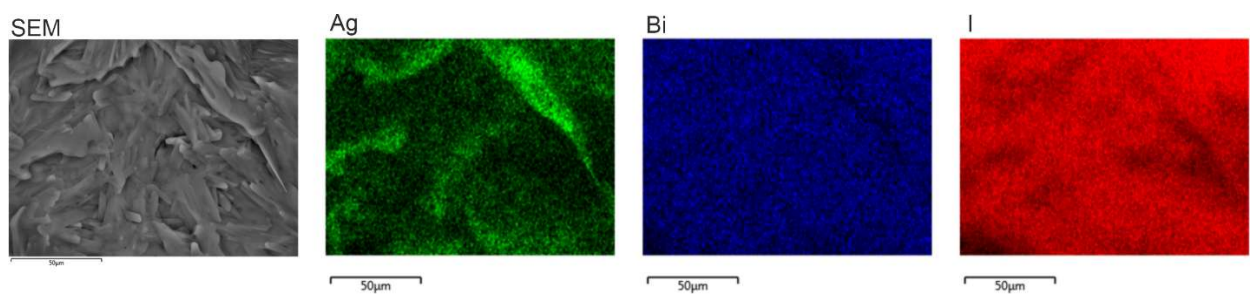

**Figure S 4.** Scanning electron microscopy overview of silver bismuth iodide synthesized using hexylamine and corresponding elemental mapping of Ag, Bi and I by energy-dispersive X-ray spectroscopy.

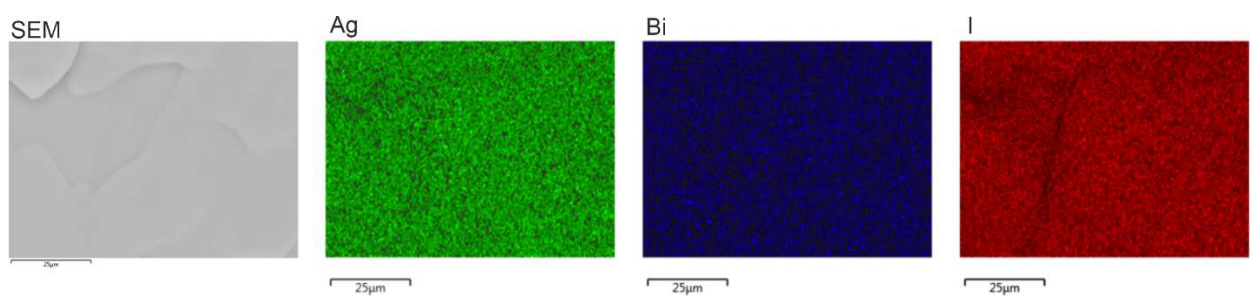

**Figure S 5.** Scanning electron microscopy overview of silver bismuth iodide synthesized using decylamine and corresponding elemental mapping of Ag, Bi and I by energy-dispersive X-ray spectroscopy.

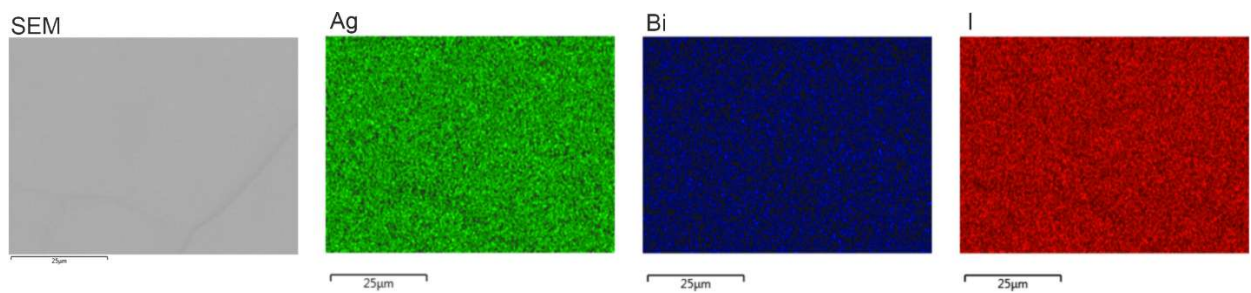

**Figure S 6.** Scanning electron microscopy overview of silver bismuth iodide synthesized using dodecylamine and corresponding elemental mapping of Ag, Bi and I by energy-dispersive X-ray spectroscopy.

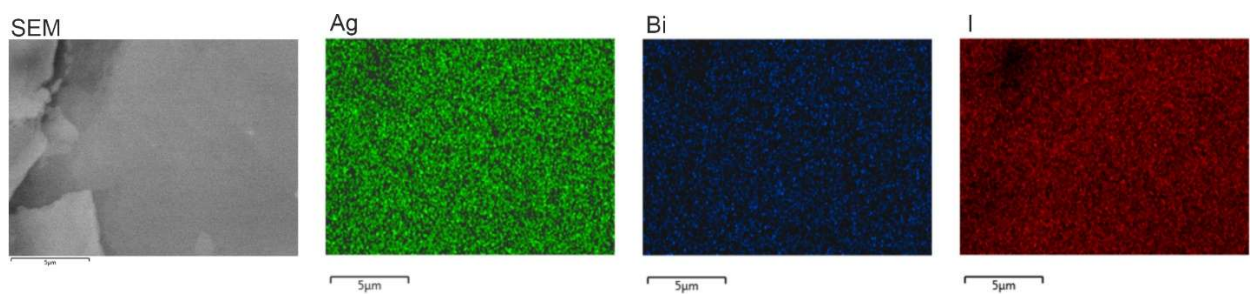

**Figure S 7.** Scanning electron microscopy overview of silver bismuth iodide synthesized using tetradecylamine and corresponding elemental mapping of Ag, Bi and I by energy-dispersive X-ray spectroscopy.

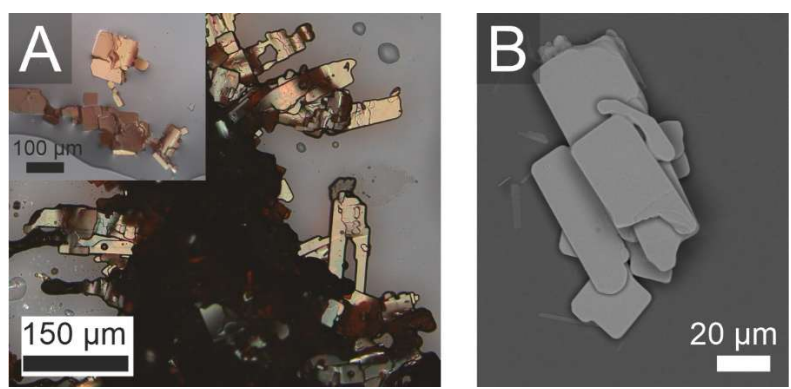

**Figure S 8.** Micrographs of recrystallized (decylamine)$_4$AgBiI$_8$ in light microscope (left) and scanning electron microscope (right), recrystallized from acetone/hexane mixture.

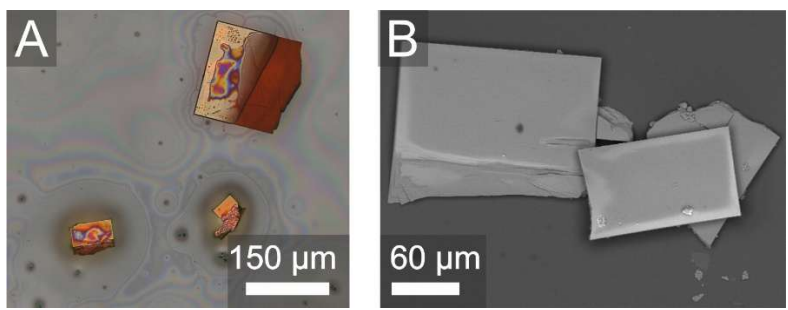

**Figure S 9.** Micrographs of recrystallized (dodecylamine)$_4$AgBiI$_8$ in light microscope (left) and scanning electron microscope (right), recrystallized from acetone/hexane mixture.

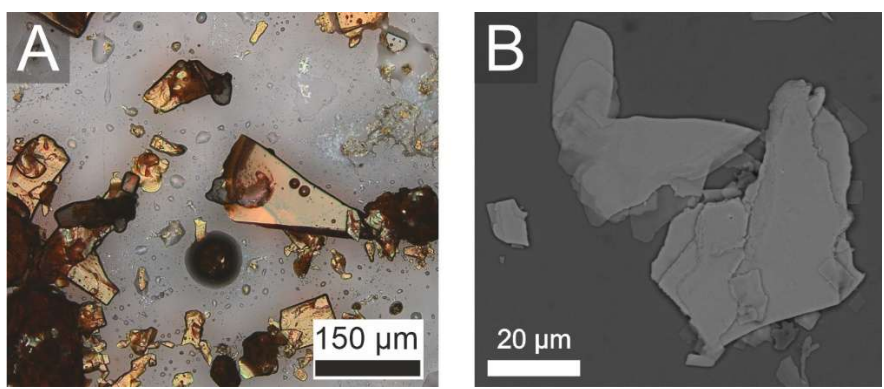

**Figure S 10.** Micrographs of recrystallized (tetradecylamine)$_4$AgBiI$_8$ in light microscope (left) and scanning electron microscope (right), recrystallized from acetone/hexane mixture.

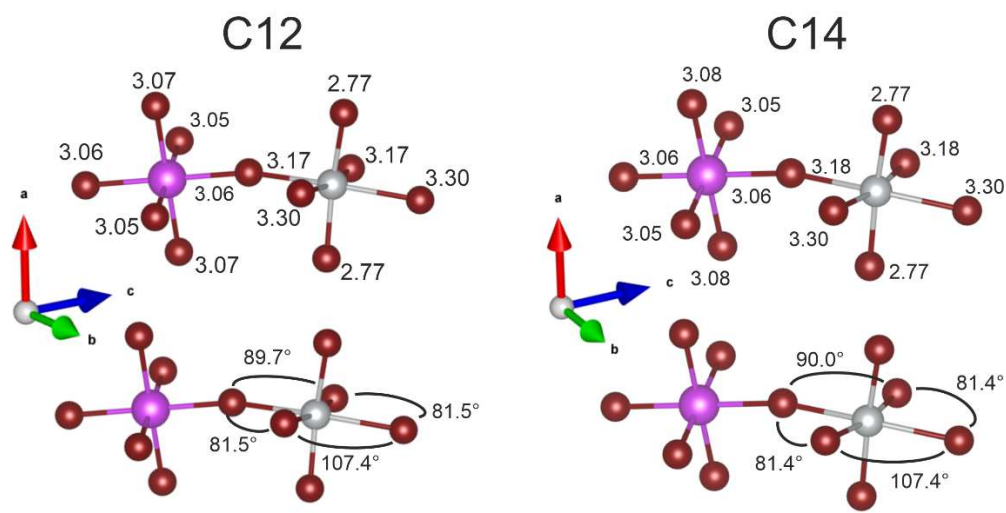

**Figure S 11.** Coordination of the metal ions in the iodide octahedra as determined by SC-XRD for dodecylammonium (left) and tetradecylammonium (right) $AgBiI_8$ with the bond lengths in Å (top) and the in-plane angles around the Ag cation (bottom).

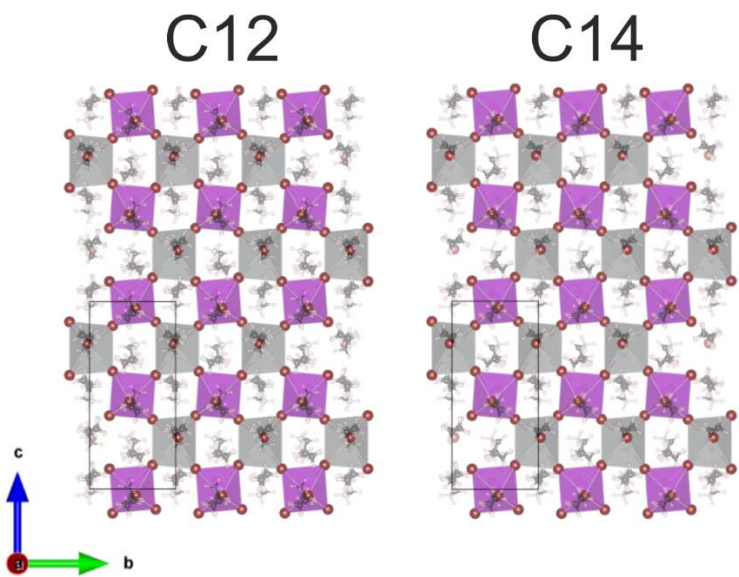

**Figure S 12**. View along the a-axis of the structures of (dodecylamine)$_4$AgBiI$_8$ (C12, left) and (tetradecylamine)$_4$AgBiI$_8$ (C14, right).

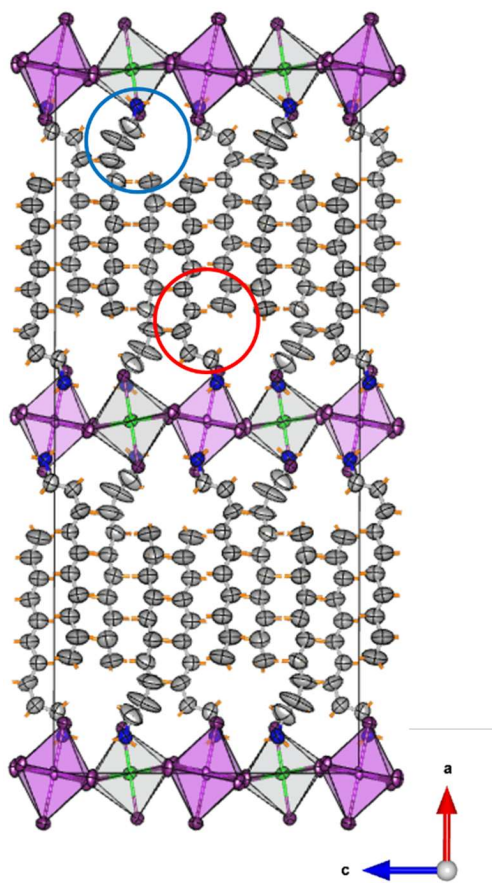

**Figure S 13.** Structure of C10 viewed along the *b*-axis with atom positions displayed as 50% probability ellipses highlighting two differently rigidly placed amines (marked blue and red). Hydrogen atoms are omitted for clarity

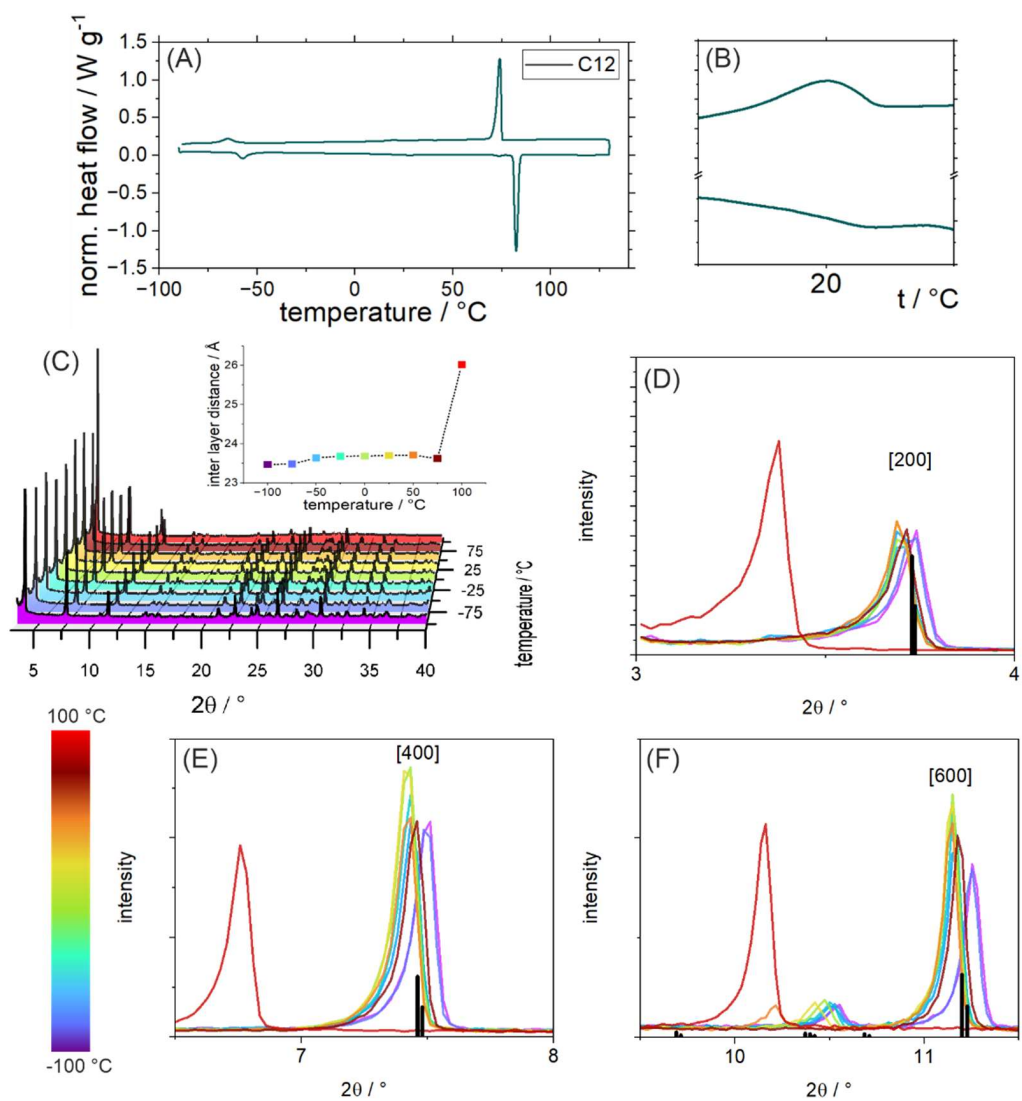

**Figure S 14.** DSC (A) of the C12 sample, the region around 20°C is shown separate (B) to visualize the small thermal event. p-XRD (C) measured at different temperatures of the C12 sample. The changes at the first 3 observed major reflections corresponding to (D) [200], (E) [400], and (F) [600] are shown in detail with the reference positions calculated from the resolved single crystal structure marked in black

Discussion of temperature dependent measurements

The temperature dependent behavior of the silver bismuth double perovskites at different temperatures was investigated by thermal analysis (differential scanning calorimetry, DSC) and temperature dependent p-XRD. In DSC, multiple reversible thermal effects are apparent around -50 °C, 0°C, and 70°C for the C10 structure (**Figure 15A,B**) and -50 °C, 20 °C and 80°C for the C12 structure (**Figure S16A,B**). The slight shift to higher transition temperatures with increasing chain length matches the chemical expectation as well as the literature reports for similar structures.[24,25] All transitions show a slight hysteresis as the transition temperature is shifting by ca. 10 °C between heating and cooling. Very similar effects can be observed in temperature dependent p-XRD as initially reflections are shifting with the temperature due to minimal changes in the lattice dimensions with temperature. A drastic change is observed above 50 °C where multiple reflections disappear while new ones emerge. This can probably be attributed to a phase transition similar to that reported for the analogue decylamine lead iodide layered perovskite in which the arrangement of the decylamine chain, i.e. the position of the ammonium and its tilt relative to the inorganic layer drastically change, resulting in a 2 Å increase in inter-layer spacing.[25] Notably, the reflections at room temperature and high temperature (100 °C) of the decylamine silver bismuth iodide layered double perovskite reported here match the trend shown for the respective reported room and high temperature decylamine lead iodide layered perovskite patterns (see **Figure S15C**). This suggests similar temperature dependent phase transitions for these two structures. When cooling below room temperature another transition is visible below -50 °C, as evidenced by a sudden shift of the reflection positions, in this case to slightly larger 2θ values, indicating a smaller inter-layer distance and therefore a denser packing of the organic layer. The event which can be seen as weak signal around 5-10 °C in the DSC is not visible in the temperature dependent diffraction. This is likely due to the similarity of the two structures. In the analogue, known[25] decylammonium PbI$_4$ perovskite the structural differences for these analogue phase transitions are found to be minimal. This correlates with the small energy difference between the two structures and the accordingly small signal in DSC. Our AgBiI$_8$ sample containing dodecylamine shows overall similar temperature dependent behavior compared to the discussed shorter chain analogue with phase transitions visible above 75 °C and below -50 °C (see **Figure S16**). As expected from the DSC measurement, the transition temperatures appear to be shifted slightly with chain length. The structural changes with a considerably larger inter-layer distance in the high temperature phase and a slight contraction in the low temperature phase also match the behavior of the shorter chain C10 sample. Again, analogues are visible to the reported dodecylammonium PbI$_4$ structure.[24] The similarity between the AgBiI$_8$ double perovskite structure reported here and the comparable PbI$_4$ perovskite reference[24,25] in their behavior with temperature is not surprising, and further supports the assumption of such linear chain amines acting as strongly structure directing compounds in the synthesis of the layered double perovskites, contrary to previous assumptions.

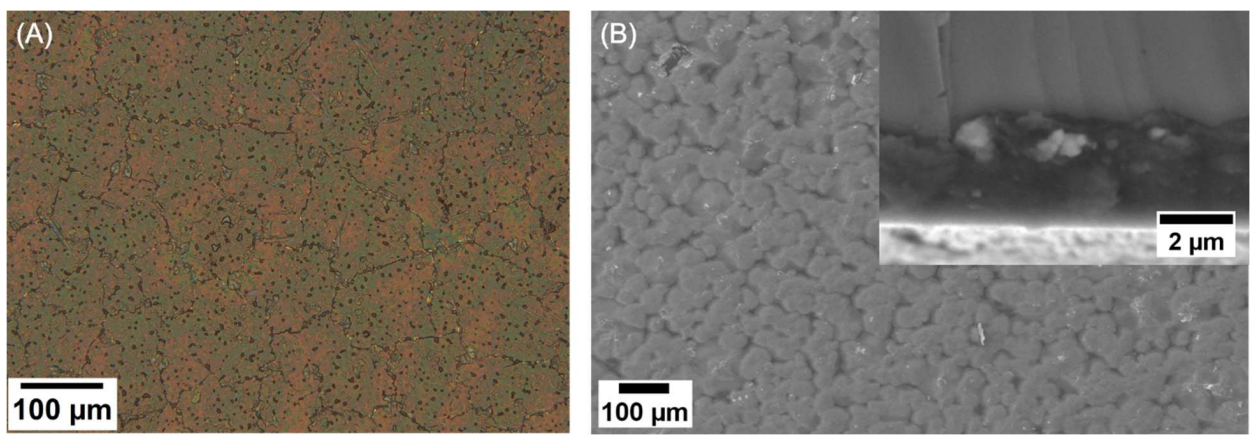

**Figure S 15.** Thin films of $(C_{12}H_{25}NH_3)_4AgBiI_8$ imaged by light microscopy (A) and scanning electron microscopy (B) in top view. The inset shows the side view of a cut film.

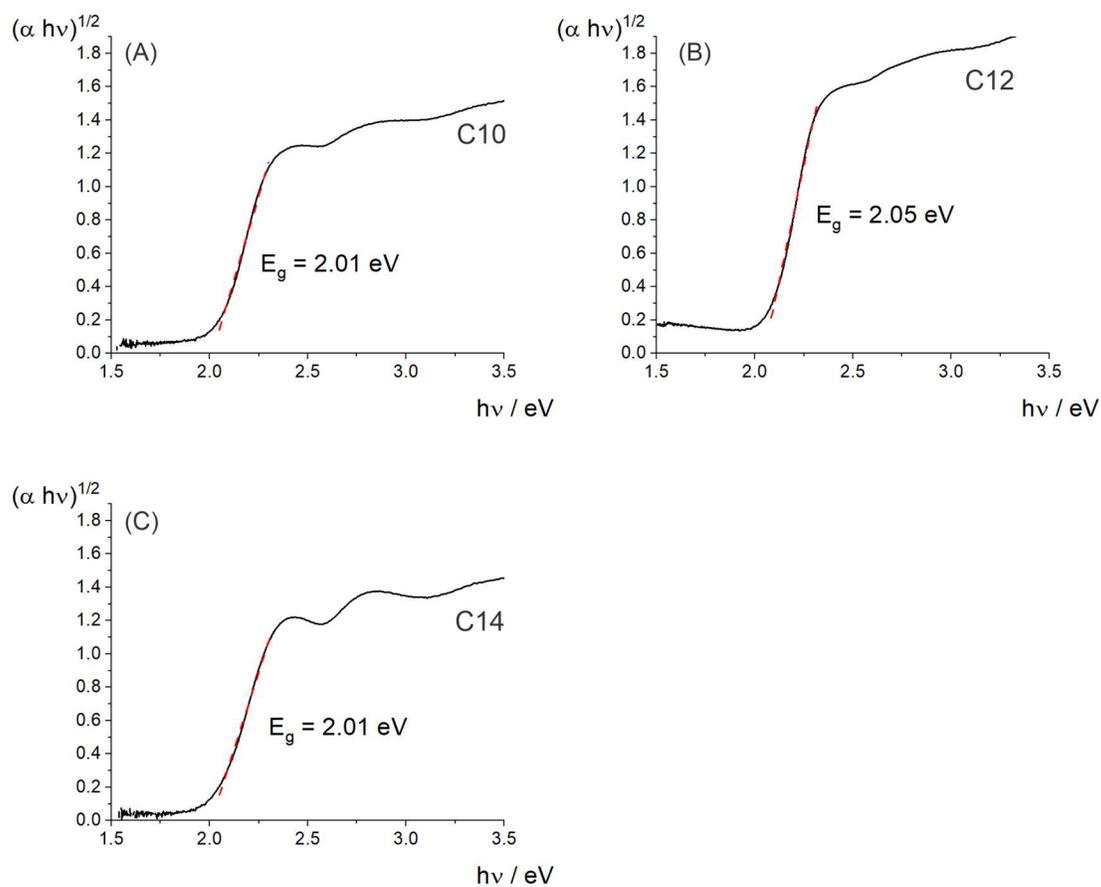

**Figure S 16.** Tauc plots of the optical spectra of $(C_{10}H_{21}NH_3)_4AgBiI_8$ (A), $(C_{12}H_{25}NH_3)_4AgBiI_8$ (B), and $(C_{14}H_{29}NH_3)_4AgBiI_8$ (C) for the case of a direct band gap with the respective band gap extracted by fitting the linear part of the plot (fitting curve shown in red dashed lines).

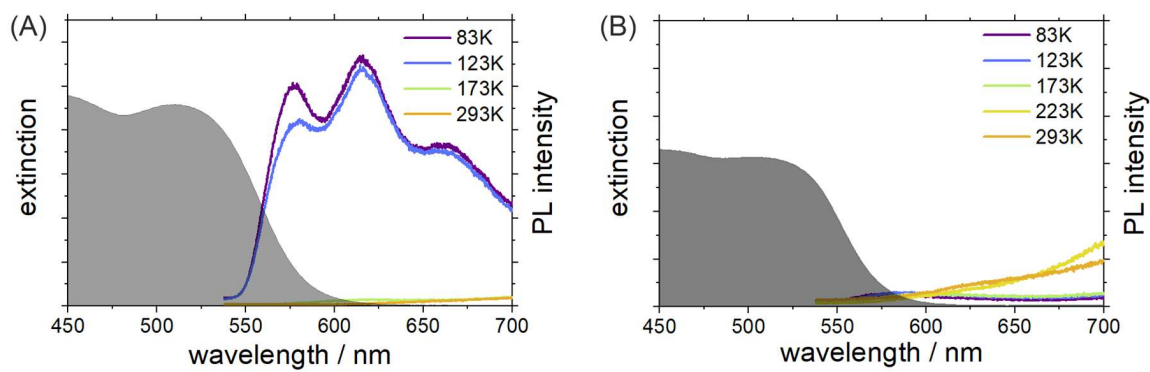

**Figure S 17**. Temperature dependent photoluminescence spectra of (A) $(C_{10}H_{21}NH_3)_4AgBiI_8$ and (B) $(C_{12}H_{25}NH_3)_4AgBiI_8$ with the corresponding absorption spectra in grey for reference.

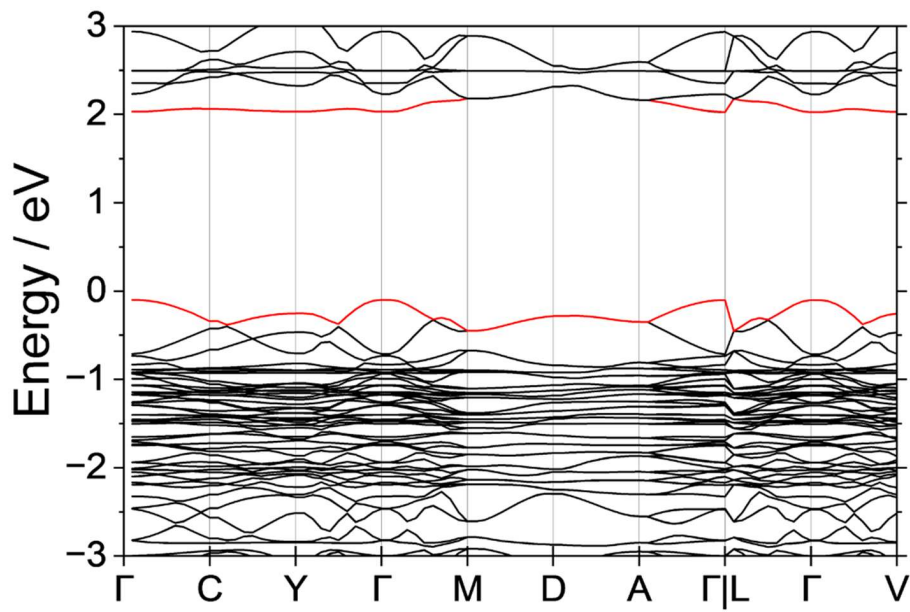

**Figure S 18.** Calculated band structure of $(C_{12}H_{25}NH_3)_4AgBiI_8$.

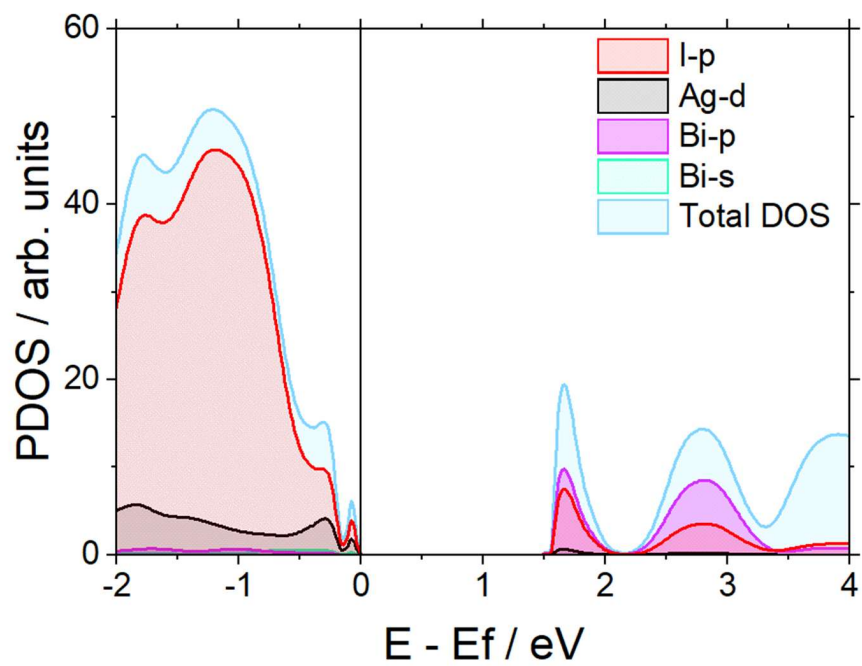

**Figure S 19.** Density of States of $(C_{12}H_{25}NH_3)_4AgBiI_8$ calculated with spin-orbit coupling.

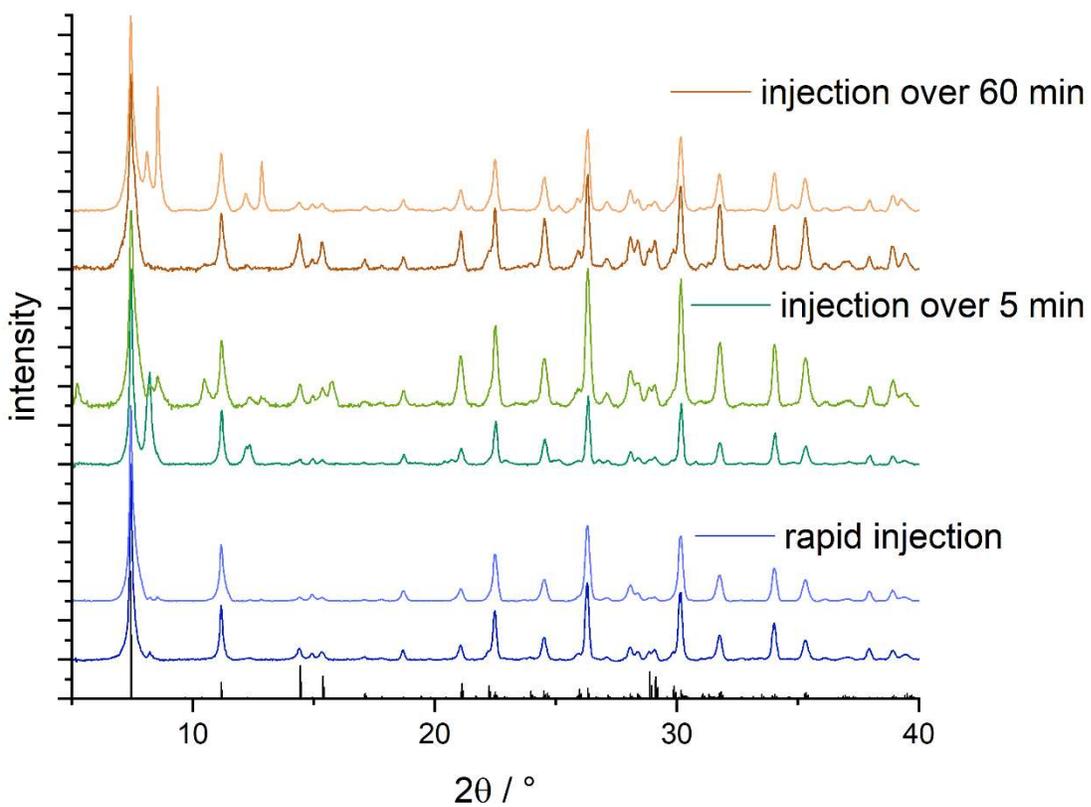

**Figure S 20.** p-XRD of C12 product obtained through rapid injection (<1 sec, blue traces) and slow addition of the iodide precursor over 5 min (green traces) and 60 min (orange traces) for two syntheses each. All products match the calculated diffraction pattern of $(C_{12}H_{25}NH_3)_4AgBiI_8$ given in black, a gradual addition leads to impurities visible by additional reflections at low 2θ.

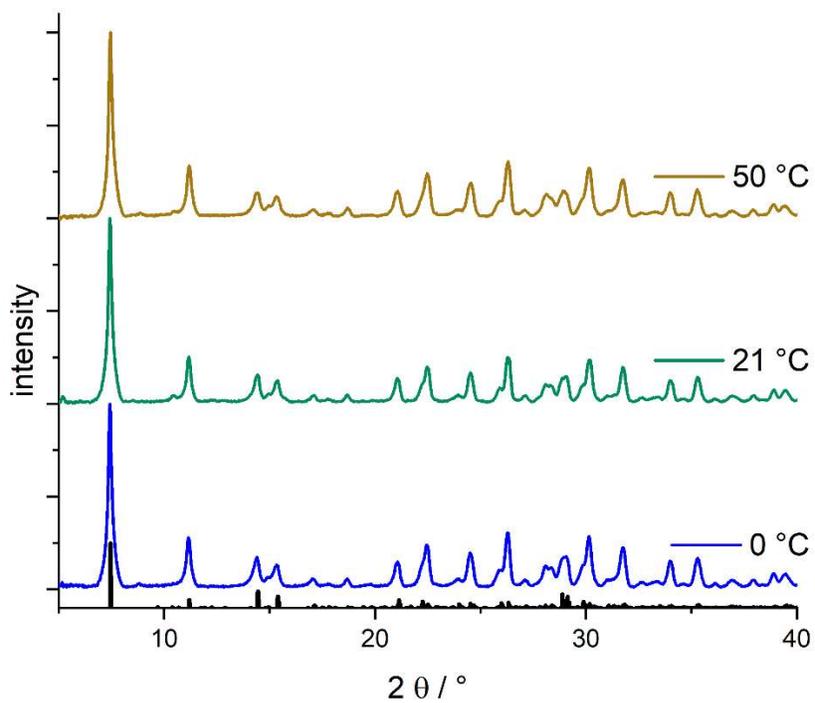

**Figure S 21.** p-XRD of C12 product obtained through rapid injection at varying temperatures. All products match the calculated diffraction pattern of $(C_{12}H_{25}NH_3)_4AgBiI_8$ given in black.

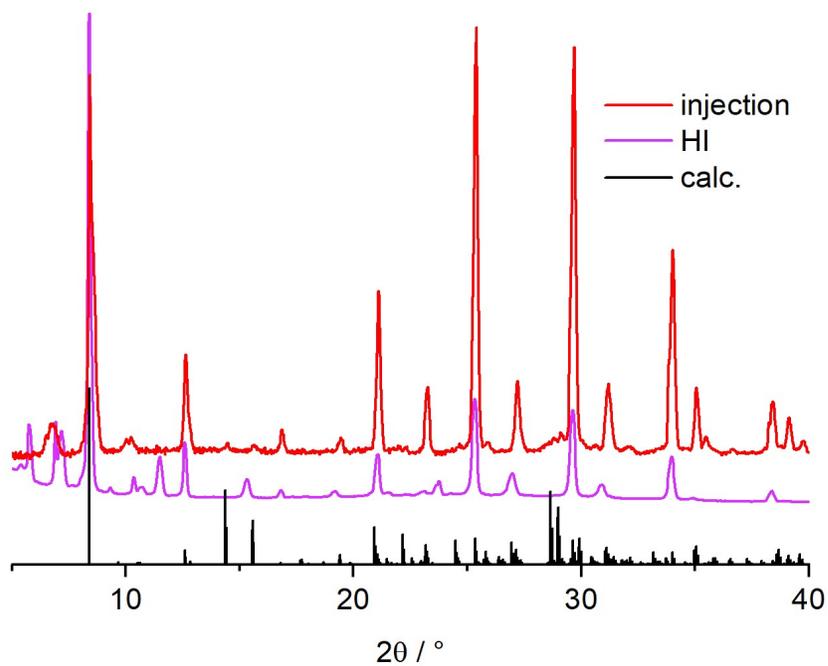

**Figure S 22.** p-XRD of drop-cast (pref. orientation) C10 product synthesized by fast precipitation after TMS-I injection (red trace) compared to a sample prepared by dissolution of 0.1 mmol AgI, 0.1 mmol BiI$_3$ and 0.4 mmol decylamine in boiling HI$_{(aq)}$ and slow cooling (purple trace). Both products match the calculated diffraction pattern of (C$_{10}$H$_{21}$NH$_3$)$_4$AgBiI$_8$ given in black, the HI synthesis has impurities visible by additional reflections at low 2θ.

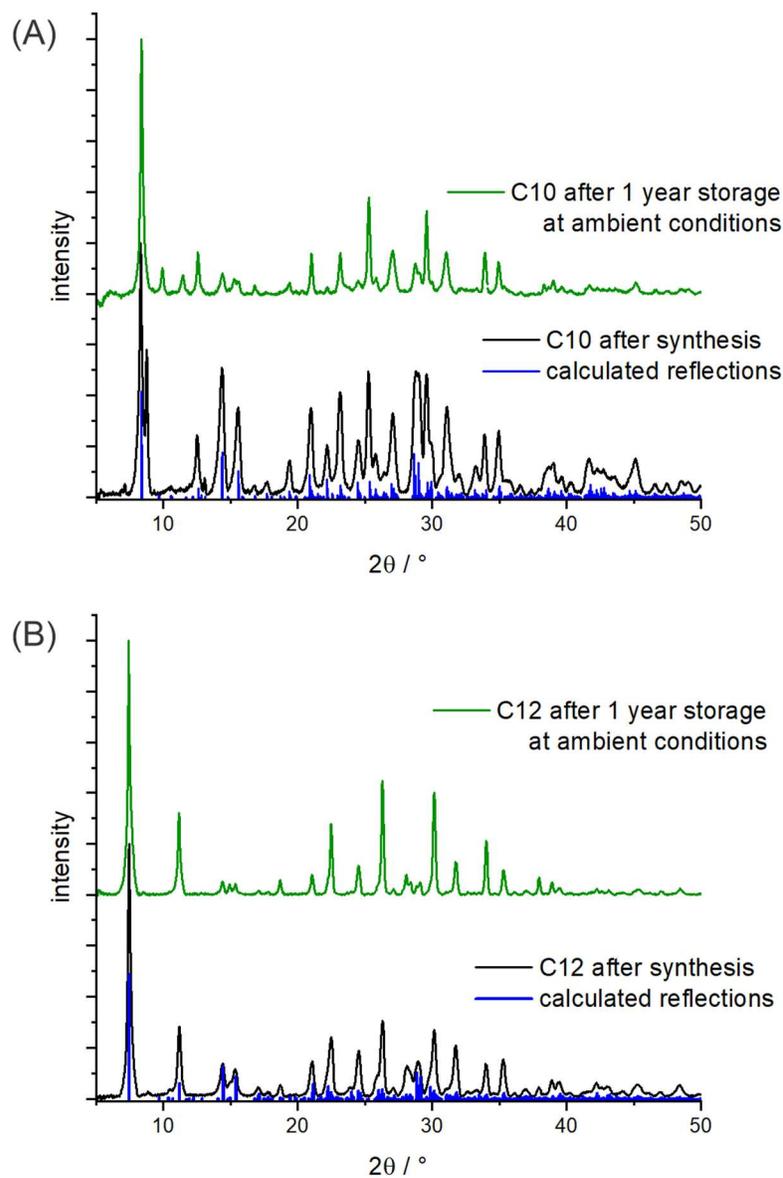

**Figure S 23.** Powder XRD of $(C_{10}H_{21}NH_3)_4AgBiI_8$ (A) and $(C_{12}H_{25}NH_3)_4AgBiI_8$ (B) after one year storage under ambient atmosphere (green traces) with comparison to the freshly synthesized powder (black traces) and the reflection positions calculated from the resolved single crystal structures (blue lines).

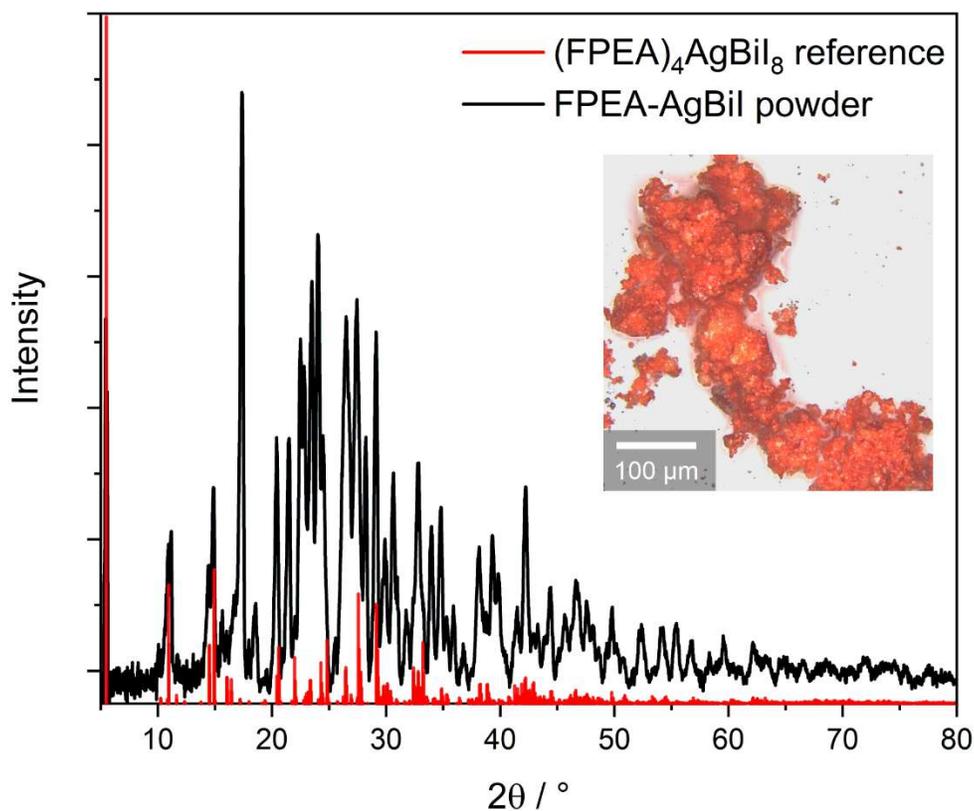

**Figure S 24**. Powder XRD of layered silver bismuth iodide synthesized by TMS-I injection using 4-fluorophenethylamine with comparison to a literature reference of this material (CCDC 2151233), which was synthesized by slow crystallization from aqueous HI.[19] The inset is showing a micrograph of the product microcrystals.

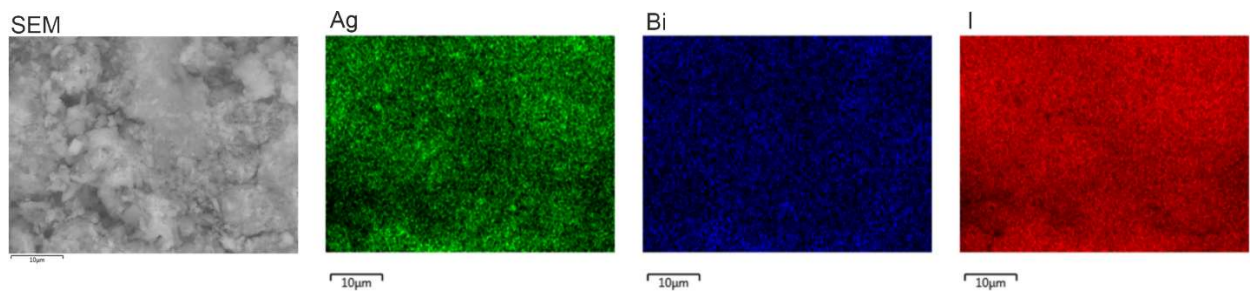

**Figure S 25.** Scanning electron microscopy overview of silver bismuth iodide synthesized using 4-fluorophenethylamine and corresponding elemental mapping of Ag, Bi and I by energy-dispersive X-ray spectroscopy (quantitative results shown in Table S2 above).

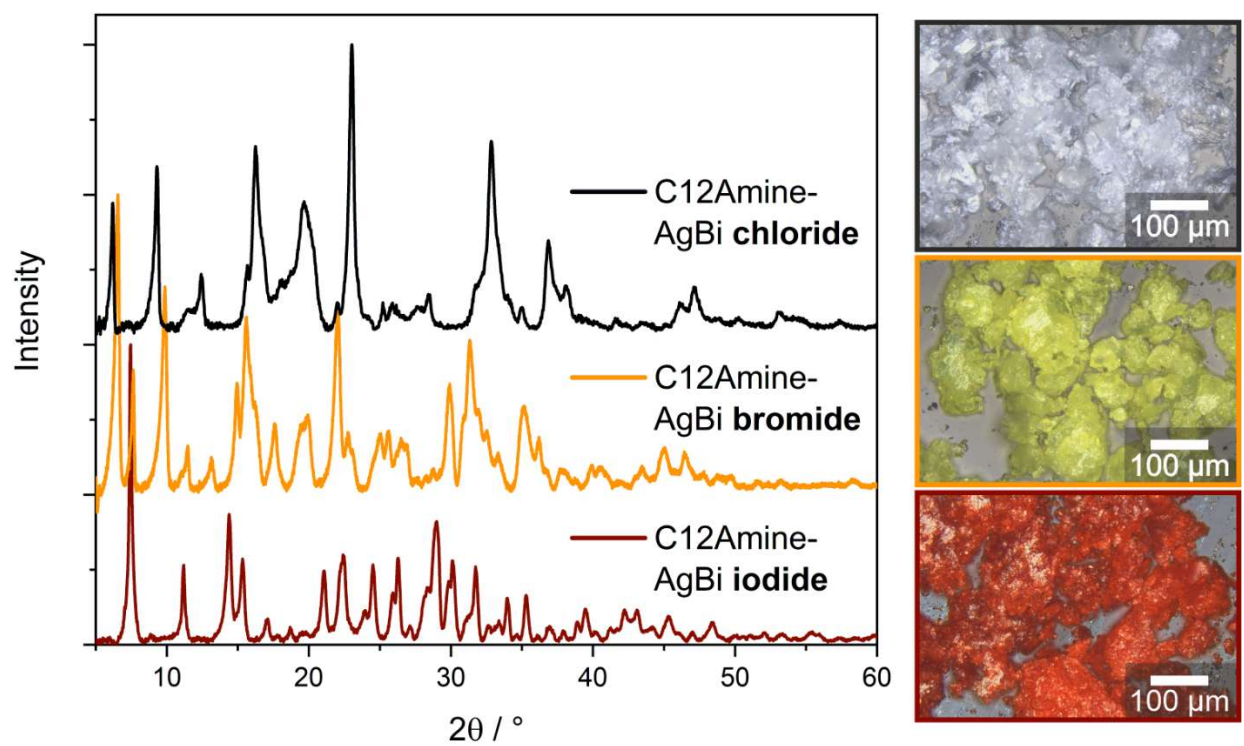

**Figure S 26.** Powder XRDs of layered silver bismuth halides (from top to bottom chloride, bromide, iodide) synthesized using dodecylamine and micrographs of the corresponding product microcrystals (in the same order top to bottom).

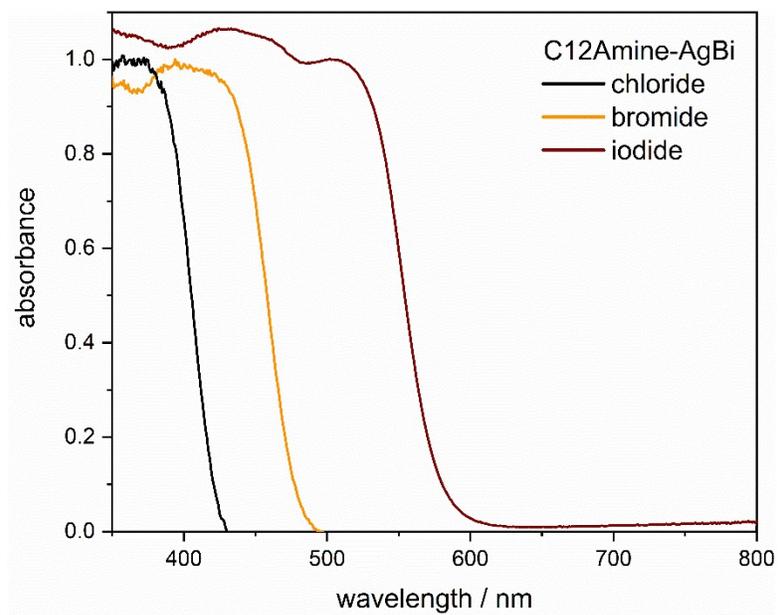

**Figure S 27.** Absorbance spectra of layered silver bismuth halides synthesized using dodecylamine.